\documentclass[nofootinbib, showkeys, twocolumn, a4paper, 10pt, aps, prd, preprintnumbers, superscriptaddress, floatfix]{revtex4-2}
\pdfoutput=1
\usepackage{amsmath,amssymb,graphicx,bm,color,slashed,array}
\usepackage[dvipsnames]{xcolor}
\usepackage[font=footnotesize,labelfont=bf,skip=5pt,justification=RaggedRight]{caption}
\usepackage{subcaption}
\usepackage{hyperref}
\hypersetup{
    colorlinks=true,
    linkcolor=blue,
    citecolor=blue,
    filecolor=magenta,
    urlcolor=cyan,
    pdftitle={The ANUBIS detector and its sensitivity to neutral long-lived particles},
    pdfpagemode=UseOutlines
    }
\usepackage{xspace}
\usepackage{enumitem}
\setlist{itemsep=0mm}

\renewcommand{\arraystretch}{0.9}

\graphicspath{{figures/}}

\newcommand{\eg}{\mbox{\itshape e.g.}\xspace}
\newcommand{\ie}{\mbox{\itshape i.e.}\xspace}

\newcommand{\proanubis}{\mbox{\textsl{pro}ANUBIS}\xspace}
\newcommand{\anubis}{\mbox{ANUBIS}\xspace}
\newcommand{\setanubis}{\mbox{\textsc{SET-ANUBIS}}\xspace}

\newcommand{\atlas}{\mbox{ATLAS}\xspace}
\newcommand{\cms}{\mbox{CMS}\xspace}
\newcommand{\mathusla}{\mbox{MATHUSLA}\xspace}
\newcommand{\codex}{\mbox{CODEX-b}\xspace}

\newcommand{\pythia}[1][]{\mbox{\textsc{Pythia#1}}\xspace}

\newcommand{\geant}{\mbox{\textsc{Geant4}}\xspace}

\newcommand{\mum}{\ensuremath{\mu\textnormal{m}}\xspace}

\newcommand{\cm}{\ensuremath{\textnormal{cm}}\xspace}
\newcommand{\metre}{\ensuremath{\textnormal{m}}\xspace}
\let\meter=\metre
\newcommand{\GeV}{\ensuremath{\textnormal{GeV}}\xspace}
\newcommand{\TeV}{\ensuremath{\textnormal{TeV}}\xspace}
\newcommand{\pb}{\ensuremath{\rm pb}\xspace}

\newcommand{\invfb}{\ensuremath{{\rm fb}^{-1}}\xspace}

\newcommand{\invab}{\ensuremath{{\rm ab}^{-1}}\xspace}

\newcommand{\stat}{\ensuremath{\textnormal{(stat)}}\xspace}
\newcommand{\syst}{\ensuremath{\textnormal{(syst)}}\xspace}

\newcommand{\met}{\ensuremath{E_T^\text{miss}}\xspace} % Missing Transverse Energy
\newcommand{\pt}{\ensuremath{p_T}\xspace} % Transverse momentum
\newcommand{\order}[1]{\ensuremath{\mathcal{O}(#1)}\xspace} % Order of X
\newcommand{\leff}{\ensuremath{L_\text{eff}}\xspace} % Effective Luminosity

\newcommand{\eps}{\varepsilon} % Efficiency
\newcommand{\acc}{\ensuremath{\mathcal A}\xspace} %Acceptance
\newcommand{\dif}{\ensuremath{{\rm d}}\xspace}

\newcommand{\dR}{\ensuremath{\Delta R}\xspace}

\newcommand{\br}{\ensuremath{\mathcal Br}\xspace}

\newcommand{\LI}{\ensuremath{\Lambda_I}\xspace}
\newcommand{\LIx}[1]{\ensuremath{\Lambda_{I,\text{#1}}}\xspace}

\def\pmbanner{}
\begin{document}
\title{\pmbanner{\LARGE \texorpdfstring{\anubis detector and its sensitivity to neutral long-lived particles\\
\includegraphics[width=0.1\linewidth]{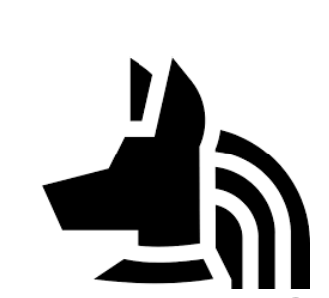}\\
\vspace{-3mm}
{\large\textbf{\anubis Collaboration}}
\\[-0.2em] \vspace{-1mm}
{\normalsize\textit{E-mail: }\href{mailto:anubis-publications@cern.ch}{anubis-publications@cern.ch}}
\\[-0.2em]
}{The \anubis detector and its sensitivity to neutral long-lived particles}}
}

\begin{abstract}
\textit{Abstract:}
Long-lived particles (LLPs), \ie particles with macroscopic lifetimes $\tau>10$~ps, appear in various extensions of the Standard Model (SM) that address fundamental questions like the particulate nature of dark matter or baryogenesis.
The \anubis detector will achieve unprecedented sensitivity to such models
compared to existing and approved experiments
by instrumenting a large decay volume adjacent to the \atlas experiment at the High-Luminosity LHC with tracking detectors.
This paper outlines the proposed detector layouts for \anubis, explores their physics potential with a scalar LLP model, and identifies the preferred layout, comparing it to other experiments.
The potential background contributions are estimated using a data-driven method, and the topology of potential background events is studied using Monte Carlo simulations.
Overall, the \anubis detector is expected to probe branching ratios down to $\order{10^{-6}}$ for exotic decays of the Higgs boson to scalar long-lived particles with masses of 10, 15, 40, and 60~\GeV and proper lifetimes of $c\tau=2.4,\, 3.0,\, 12$, and 18~\metre, respectively.
Moreover, for branching ratios of 0.1\% of the Higgs boson into long-lived scalars with a mass of 15~\GeV, the \anubis detector can probe a $c\tau$ range between $1.1\times10^{-1}~\metre$ and $4.0\times10^3~\metre$.
\end{abstract}

% authors (by surname)
%
% Re-encoded from the authblk form of ceil/ceil/authorList.tex into the revtex
% superscriptaddress form used by hnl/hnl/authorList.tex. With
% superscriptaddress, identical \affiliation strings are collapsed and
% numbered automatically, so the explicit \affil[n] indices are gone: each
% author simply carries its affiliations, in the order they had in the old
% numbering. The strings themselves are unchanged apart from "Th\'eo" taking
% the accent it has in the HNL and ALP papers.
%
\author{Thomas Adolphus}
\affiliation{Formerly at: Cavendish Laboratory, University of Cambridge, Cambridge, United Kingdom}
\author{Giulio Aielli}
\affiliation{University and INFN Roma Tor Vergata, Rome, Italy}
\author{Jahanzeb Akhtar}
\affiliation{School of Sciences and Engineering, Glasgow Caledonian University, Glasgow, United Kingdom}
\affiliation{Formerly at: Cavendish Laboratory, University of Cambridge, Cambridge, United Kingdom}
\author{Arun Atwal}
\affiliation{Formerly at: Cavendish Laboratory, University of Cambridge, Cambridge, United Kingdom}
\author{Martin Bauer}
\affiliation{IPPP, Dept of Physics, Durham University, Durham, United Kingdom}
\author{Rachel Bentham}
\affiliation{Department of Physics and Astronomy, University of Manchester, Manchester, United Kingdom}
\affiliation{Formerly at: Cavendish Laboratory, University of Cambridge, Cambridge, United Kingdom}
\author{Oleg Brandt}
\affiliation{Cavendish Laboratory, University of Cambridge, Cambridge, United Kingdom}
\author{Jude Burling}
\affiliation{Formerly at: Cavendish Laboratory, University of Cambridge, Cambridge, United Kingdom}
\author{Jon Burr}
\affiliation{Formerly at: Cavendish Laboratory, University of Cambridge, Cambridge, United Kingdom}
\author{Patrick Collins}
\affiliation{Formerly at: Cavendish Laboratory, University of Cambridge, Cambridge, United Kingdom}
\author{Louie Dartmoor Corpe}
\affiliation{Universit\'e Clermont Auvergne / Laboratoire de Physique de Clermont Auvergne, CNRS/IN2P3, France}
\author{Matthew Coxon}
\affiliation{Formerly at: Cavendish Laboratory, University of Cambridge, Cambridge, United Kingdom}
\author{Jonas Dej}
\affiliation{Cavendish Laboratory, University of Cambridge, Cambridge, United Kingdom}
\author{Sofie Nordahl Erner}
\affiliation{Formerly at: IPPP, Dept of Physics, Durham University, Durham, United Kingdom}
\author{Cayetano Fernandez Ruiz}
\affiliation{Cavendish Laboratory, University of Cambridge, Cambridge, United Kingdom}
\author{Jindrich Jelinek}
\affiliation{Institute of Experimental and Applied Physics, CTU in Prague, Czech Republic}
\affiliation{Department of Nuclear and Particle Physics, University of Geneva, Geneva, Switzerland}
\affiliation{Formerly at: Cavendish Laboratory, University of Cambridge, Cambridge, United Kingdom}
\author{Oliver Kortner}
\affiliation{Max Planck Institute for Physics, Munich, Germany}
\author{Hubert Kroha}
\affiliation{Max Planck Institute for Physics, Munich, Germany}
\author{Lawrence Lee Jr}
\affiliation{Department of Physics, University of Tennessee, Knoxville, TN 37996, USA}
\affiliation{Formerly at: Department of Physics, Harvard University, Cambridge, USA}
\author{Christopher Lester}
\affiliation{Cavendish Laboratory, University of Cambridge, Cambridge, United Kingdom}
\author{Yingshan Liang}
\affiliation{Cavendish Laboratory, University of Cambridge, Cambridge, United Kingdom}
\author{Kaijia Liu}
\affiliation{Formerly at: Cavendish Laboratory, University of Cambridge, Cambridge, United Kingdom}
\author{Anna Mullin}
\affiliation{Cavendish Laboratory, University of Cambridge, Cambridge, United Kingdom}
\author{Christian Ohm}
\affiliation{KTH Royal Institute of Technology, Stockholm, Sweden}
\author{David Peng}
\affiliation{Formerly at: Cavendish Laboratory, University of Cambridge, Cambridge, United Kingdom}
\author{Luca Pizzimento}
\affiliation{Department of Physics, University of Hong Kong, Hong Kong SAR, China}
\author{Ludovico Pontecorvo}
\affiliation{CERN, Geneva, Switzerland}
\author{Giorgia Proto}
\affiliation{Max Planck Institute for Physics, Munich, Germany}
\author{Th\'eo Reymermier}
\affiliation{Institut de Physique des 2 Infinis (IP2I), Lyon, France}
\affiliation{Formerly at: Cavendish Laboratory, University of Cambridge, Cambridge, United Kingdom}
\author{Michael Revering}
\affiliation{Cavendish Laboratory, University of Cambridge, Cambridge, United Kingdom}
\author{Elisa Ritz}
\affiliation{Formerly at: Hamburg University, Hamburg, Germany}
\author{Thomas P. Satterthwaite}
\affiliation{Department of Physics, Stanford University, Stanford, CA 94305, USA}
\affiliation{Kavli Institute for Particle Astrophysics and Cosmology, Stanford, CA 94305, USA}
\affiliation{Formerly at: Cavendish Laboratory, University of Cambridge, Cambridge, United Kingdom}
\author{Aashaq Shah}
\affiliation{Cavendish Laboratory, University of Cambridge, Cambridge, United Kingdom}
\author{Sinem Simsek}
\affiliation{Istinye University, Istanbul, Turkey}
\author{Daniel Soyk}
\affiliation{Max Planck Institute for Physics, Munich, Germany}
\author{Tom Spencer}
\affiliation{School of Sciences and Engineering, Glasgow Caledonian University, Glasgow, United Kingdom}
\affiliation{Formerly at: Cavendish Laboratory, University of Cambridge, Cambridge, United Kingdom}
\author{Paul Swallow}
\affiliation{Cavendish Laboratory, University of Cambridge, Cambridge, United Kingdom}
\author{Balint Szepfalvi}
\affiliation{Formerly at: Cavendish Laboratory, University of Cambridge, Cambridge, United Kingdom}
\author{Noshin Tarannum}
\affiliation{Department of Nuclear and Particle Physics, University of Geneva, Geneva, Switzerland}
\affiliation{Formerly at: Cavendish Laboratory, University of Cambridge, Cambridge, United Kingdom}
\author{Olivia Valentino}
\affiliation{Imperial College London, London, United Kingdom}
\affiliation{Formerly at: Cavendish Laboratory, University of Cambridge, Cambridge, United Kingdom}
\author{Julian Wack}
\affiliation{Cavendish Laboratory, University of Cambridge, Cambridge, United Kingdom}
\author{Yanglin Wan}
\affiliation{School of Physics and Astronomy, University of Edinburgh, Edinburgh, United Kingdom}
\affiliation{Formerly at: Cavendish Laboratory, University of Cambridge, Cambridge, United Kingdom}
\author{Peng Wang}
\affiliation{Formerly at: Cavendish Laboratory, University of Cambridge, Cambridge, United Kingdom}
\author{Esperanza Winter Lopez}
\affiliation{Formerly at: Cavendish Laboratory, University of Cambridge, Cambridge, United Kingdom}
\author{Monami Yoshioka}
\affiliation{Formerly at: Cavendish Laboratory, University of Cambridge, Cambridge, United Kingdom}
\author{and Yingchang Zhang}
\affiliation{Cavendish Laboratory, University of Cambridge, Cambridge, United Kingdom}
\author{for the ANUBIS Collaboration}
\noaffiliation


\def\andname{\hspace*{-0.5em}}
\maketitle

\hbox{}
\clearpage

%%%%%%%%%%%%%%%%%%%%%%%%%%%%%%%%%%%%%%%
\section{Introduction} \label{sec:Intro}

Particles with macroscopic lifetimes $\tau>10$~ps, also known as long-lived particles~(LLPs), are a key feature of many Beyond the Standard Model~(BSM) theories, including supersymmetry~\cite{Barbier:2004ez,Giudice:1998bp}, hidden sector models~\cite{Han:2007ae,Strassler:2006im}, and exotic Higgs decays~\cite{Chacko:2005pe,Cai:2008au}. 
These particles can provide insights into problems like naturalness~\cite{Giudice:1998bp,Burdman:2006tz}, dark matter~\cite{Hall:2010jx,Cheung:2010gk,Zurek:2013wia}, and baryogenesis~\cite{Barry:2013nva}. 
Their lifetime, $\tau$, is typically a free model parameter and hence can vary widely, from $\tau>10$~ns and up to the Big Bang nucleosynthesis limit of $\mathcal{O}$(0.1 s)~\cite{Fradette:2017sdd}.
%for an exotic Higgs boson decay to BSM scalar particles. 
This offers considerable flexibility for LLP candidates, making them a versatile but challenging target in the search for new physics.

Detecting LLPs, especially those that are neutral and have long lifetimes, poses significant experimental challenges. The general purpose detectors at the Large Hadron Collider (LHC)~\cite{Evans:1129806} such as \atlas~\cite{Aad:1129811}, \cms~\cite{Chatrchyan:1129810}, or LHCb~\cite{LHCb:2008vvz} are limited by their finite detection volumes, intricate backgrounds, and trigger requirements that were designed for prompt particles~\cite{Alimena:2019zri, CMS:2021sch, CMS:2021juv,ATLAS:2018tup}. 
Despite a concerted experimental effort from \atlas, \cms, and LHCb, particles with long lifetimes that decay outside the active detector volume can escape detection.

\begin{figure}[htpb]
    \centering
    \includegraphics[width=0.95\linewidth]{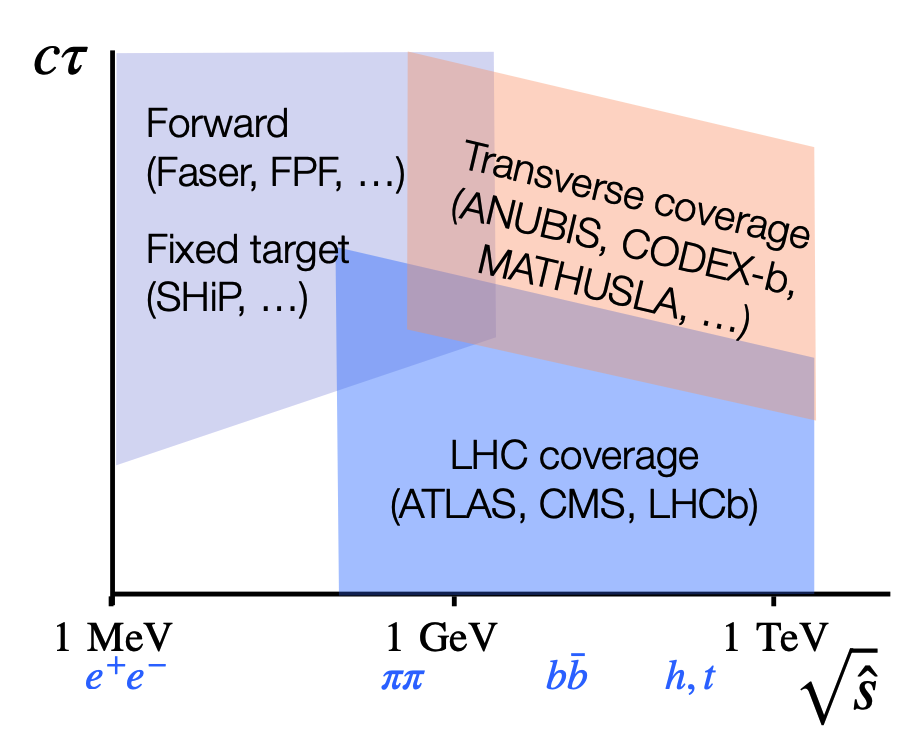}
    \caption{A schematic representation of the coverage of different experiments to the proper lifetime $c\tau$, and partonic centre-of-mass energy, $\sqrt{\hat{s}}$.}
    \label{fig:LLP_Complimentarity}
\end{figure}

To extend the sensitivity to LLPs with longer lifetimes, several proposals for dedicated detectors have emerged. 
These can be roughly split into two categories that complement general purpose detectors: {\em forward detectors} that are positioned along the beamline 
and {\em transverse detectors} that are positioned transverse to the beamline close to an LHC interaction point.
The complementarity in LLP parameter space coverage between general purpose, transverse, and forward detectors is highlighted by Physics Beyond Colliders (PBC) initiative~\cite{PBC:2025sny} and illustrated in Figure~\ref{fig:LLP_Complimentarity}.

While forward detectors such as FASER~\cite{Antel:2023hkf} can effectively probe LLP scenarios that can be produced at small partonic centre-of-mass energies, $\sqrt{\hat s}$, of $\order{1~\GeV}$, their ability to search for scenarios with $\sqrt{\hat s}\gtrsim10~\GeV$ is kinematically restricted. 
The same restriction also applies to beam dump experiments such as SHiP~\cite{Ahdida:2654870}.
By contrast, models where the LLPs are produced at the electroweak scale of $\sqrt{\hat s}\approx 100~\GeV$ and above can be effectively probed with transverse experiments~\cite{bib:anubis_orig,Gligorov:2018vkc,Gligorov:2017nwh,Chou:2016lxi,Curtin:2018mvb, MATHUSLA:2020uve}, but are kinematically inaccessible for forward experiments. 
Hence, transverse and forward detectors cover unique areas of LLP phase space, as shown in Figure~\ref{fig:LLP_Complimentarity}, but there is also an overlap in experimental coverage which allows for comparison between them.

This paper presents the physics case for the \anubis (AN Underground Belayed In-Shaft) detector, which proposes the installation of an array of tracking detectors above the \atlas experiment at CERN~\cite{bib:anubis_orig}. 
By leveraging the existing infrastructure, \anubis aims to search for LLPs with proper lifetimes $c\tau$ ranging from $\order{1\text{ m}}$ to $\mathcal{O}(10^5\text{ m})$, complementing the sensitivity of \atlas. 
It offers a cost-effective means of enhancing LLP searches, by modifying the \atlas experimental environment. 

A prototype demonstrator \proanubis consisting of one \anubis tracking station module was installed in 2023 close to the ceiling of the \atlas cavern~\cite{Shah:2024fpl,ANUBIS:2025mme} and successfully commissioned~\cite{ANUBIS:2025inn}. 
Since then, the \proanubis detector has recorded 104~\invfb and 73~\invfb of $pp$ collision data at $\sqrt s = 13.6~\TeV$ in 2024 and 2025, respectively. 
The main physics goal of \proanubis is a direct measurement of hadron interactions in a background-enriched region, which then can be used to refine the tracking and vertexing algorithms, inform detector performance benchmarks for \anubis, improve background simulations, and further optimisations.
First \proanubis performance results were published recently~\cite{ANUBIS:2026jxg}, and the deeper analysis of \proanubis data is underway. 

This document is structured as follows: Section~\ref{sec:ANUBIS_overview} presents the detector geometry configurations that are considered for \anubis, followed by a detailed discussion of the relevant background sources predictions, an evaluation of their potential contributions, and a study of their topological signature in Section~\ref{sec:bkg}. 
Finally, a study of the projected sensitivity to LLPs using a standard benchmark endorsed by the Physics Beyond Colliders initiative~\cite{Beacham:2019nyx} of CERN is presented in Section~\ref{sec:SensitivityStudies}.
Based on these findings, a strong case is made for the construction of the \anubis detector.

%========================================
\section{The ANUBIS project and its layout} \label{sec:ANUBIS_overview}
\begin{figure}
\centering
\includegraphics[height=10.5cm]{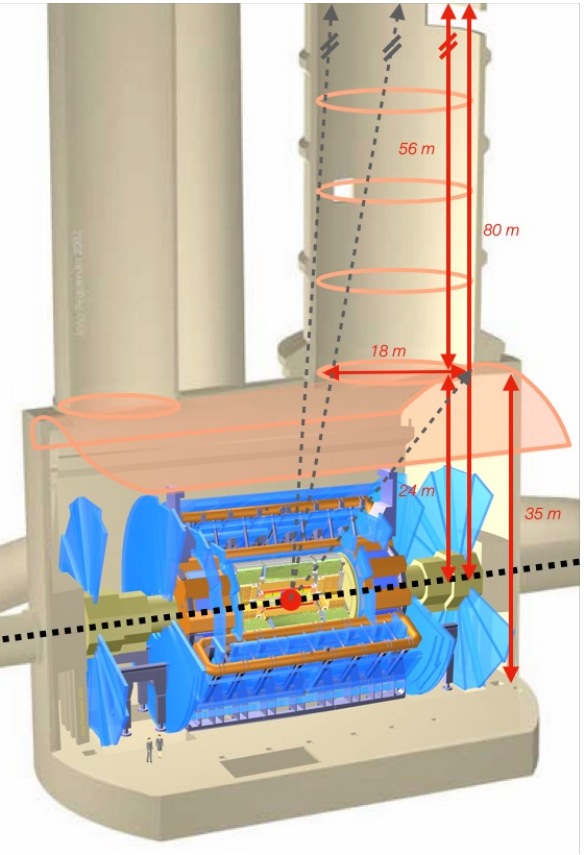} 
\caption{The layout of the underground cavern at LHC Point 1, UX15, featuring the \atlas experiment and the PX14 and PX16 service shafts above it. The two proposed configurations of the \anubis detector are shown in orange, where the shaded area illustrates the potential ceiling configuration. This includes two circular stations at the bottom of the two access shafts. The shaft configuration is indicated by the four circular stations within the PX14 service shaft, and has a solid orange line.} 
\label{fig:SketchATLAS_undergroundCavern}
\end{figure}

The \anubis detector exploits the `smoking gun' signature of neutral LLP decays: an isolated displaced vertex (DV) of charged particle tracks emanating from the decay position of the LLP some distance away from the interaction point. 
It will utilise Resistive Plate Chambers (RPCs) as the tracking technology. A triplet of RPCs stacked on top of each other form a `tracking layer', and two tracking layers separated by $\sim1$~m compose a Tracking Station (TS) of \anubis~\cite{bib:anubis_orig}.
Two detector configurations of the \anubis project, as depicted in Figure~\ref{fig:SketchATLAS_undergroundCavern}, are evaluated through simulation studies. 

The `shaft configuration' was the first studied and involves suspending four tracking stations within the \atlas PX14 service shaft, at distances ranging from 23 to 80 meters from the beamline. 
This corresponds to the layout in the original \anubis proposal~\cite{bib:anubis_orig}. 
However, this geometry is subject to operational complications that became apparent in discussions with the \atlas technical coordination team~\cite{bib:ludo}. 
For example, the TSs of \anubis would require removal during periods such as a Year-End Technical Stop (YETS) when the service shafts are required for crane access to the \atlas experimental cavern, UX15. 
While this does not represent an insurmountable challenge, it is considered impractical given the time required to extract four TSs and the limitations on space in the SX15 hall above the PX14 service shaft that is needed for the disassembly of the TSs for further transport and storage during the YETS.
Moreover, the Point 1 experimental site of the \atlas detector does not provide sufficient space to host four TSs, which would require an involved transport to the Meyrin site, where space is also limited.
Besides these two points, the discussions with the \atlas technical coordination team and with CERN civil engineering experts did not highlight any other critical points for the installation of the \anubis detector.

The other configuration considered is the `ceiling configuration' where two tracking layers separated by 1 m are installed directly onto the ceiling of the \atlas cavern, approximately 23 m above the beamline, alongside two circular stations that are positioned at the bottom of the PX14 and PX16 service shafts. 
As in the original `shaft configuration', these two TSs at the bottom of the two service shafts would also need to be removed when they are required by \atlas. 
These TSs can be extracted as several parts rather than a full circular station, but having only one TS per shaft will reduce the space requirement in the limited temporary storage/disassembly location immediately next to the shaft in the SX1 hall above the UX15 \atlas cavern, and on the mid-term storage for the TSs during YETS.
Moreover, this configuration would have a larger solid angle coverage than the shaft configuration, offering a broader detection range, making it highly effective for capturing LLP decays that fall between the \atlas vertex detector's limits and the cavern ceiling, but this also leads to a smaller decay length acceptance.
Hence, the ceiling configuration is considered the default in the following.

It is foreseen to fully integrate \anubis with \atlas, which will enable \anubis to initiate trigger signals for \atlas and vice versa. 
Technically, this is possible given that the \atlas Level 0 trigger at the High-Luminosity LHC (HL-LHC) will operate with a latency window of not less than 5.5~$\mu\text{s}$~\cite{bib:sankey}, while \anubis' latency is expected to be less than 3~$\mu\text{s}$. This is mostly due to the signal propagation time and the data acquisition electronics. Both tracking and vertexing are computationally straightforward since the stray magnetic field of \atlas should not affect the trajectories of charged particles outside of \atlas significantly~\cite{ATLAS:1997ad}.
The full integration with \atlas will benefit \anubis in several ways:
\begin{itemize}
\item
First and foremost, the full record of a $pp$ collision with a DV from a candidate LLP event is essential for reducing background contributions, as will be discussed in Section~\ref{sec:bkg}. 
For example, a true DV detected by \anubis is likely to coincide azimuthally with significant missing transverse momentum (\met) measured by \atlas\footnote{%
\anubis adopts the right-handed coordinate system of \atlas, where its origin is at the nominal interaction point in the centre of the \atlas detector and the $z$-axis points along the beam pipe. The $x$-axis points from the Interaction Point (IP) to the centre of the LHC ring, and the $y$-axis points upwards.
Polar coordinates ($r$, $\phi$) are used in the transverse plane, $\phi$ being the azimuthal angle around the $z$-axis. 
The pseudorapidity is defined in terms of the polar angle $\theta$ as $\eta \equiv-\ln \tan(\theta/2)$ and is equal to the rapidity $y = \frac12 \ln \left(\frac{E+p_z}{E-p_z}\right)$
in the relativistic limit.
Angular distance is measured in units of $\Delta R \equiv \sqrt{\Delta\eta^2+\Delta\phi^2}$.%
}.
\item
Second, a {\em continuous} active decay volume would be established between \atlas and \anubis, extending from the interaction point to approximately the ceiling of the \atlas cavern.
This maximises the combined sensitivity of \atlas and \anubis to LLP scenarios with decay lengths $L$ ranging from a few $\mum$ to tens of metres.
\item
Third, the full integration allows \atlas and \anubis to probe scenarios where LLPs are produced in association with prompt SM particles like \eg $W$ or $Z$ bosons in scenarios with heavy neutral leptons~\cite{Brdar_2019,Bondarenko:2018ptm,Gorbunov:2007ak} or 
%photophobic axion-like particle models
models with exotic Higgs decays into axions such as $h \to Za$ that could provide unique information on the UV completion of the axion theory~\cite{ATLAS:2025pak,Bauer:2017ris}. These are expected to have negligible background contributions.
\end{itemize}

%%%%%%%%%%%%%%%%%%%%%%%%%%%%%%%%%%%%%%%%%%%%%%
\section{Backgrounds}
\label{sec:bkg}

After reviewing potential sources of backgrounds for the \anubis detector in Section~\ref{sec:bkg_sources}, Section~\ref{sec:bkg_reduction} discusses the reduction of these backgrounds by leveraging the passive shielding effect of the \atlas detector and the information about the prompt part of the $pp$ collision event recorded by the \atlas detector.
This strategy provides an almost background-free experimental environment for LLP searches with the \anubis detector, as demonstrated using a data-driven method in Section~\ref{sec:bkg_predictions}.
The experimental signature of hadronic interactions that represent the dominant background contribution is studied using simulations in Section~\ref{sec:bkg_signatures}.

%%%%%%%%%%%%%%%%%%%%%%%
\subsection{Background sources}
\label{sec:bkg_sources}
Several types of background sources can potentially contribute to LLP detectors like \anubis:
\begin{itemize}[itemsep=0em]
\item
cosmic rays;
\item
beam-induced backgrounds like beam-gas and beam-collimator interactions;
\item
backgrounds from decays of quasi-thermal neutrons;
\item
random coincidences of unrelated tracks;
\item
from SM particles with macroscopic lifetimes, subsequently referred to as `SM LLPs'.

\end{itemize}

The first three background sources are evenly distributed in time and can be effectively rejected using the excellent timing resolution of \anubis or the event topology, as discussed in the original \anubis proposal~\cite{bib:anubis_orig}. 
The background from random coincidences of unrelated tracks is inversely proportional to the distance from the interaction point and is negligible for ${L\gtrsim 1~\meter}$, which is also discussed in Ref.~\cite{bib:anubis_orig}.

The dominant potential background contribution to \anubis is from SM LLPs. There are only two electrically neutral SM LLPs that are relevant for \anubis given its large active decay volume several metres away from the interaction point: long-lived neutral kaons, $K_L^0$, and neutrons, $n$. 
These are characterised by mean proper lifetimes of $c\tau=15.3$~m and $2.7\times10^{11}$~m, respectively~\cite{bib:pdg2024}.
SM LLPs can contribute via two principal mechanisms: decays into charged particles that are registered as a DV, and the production of hadrons in collisions with material, which shall be referred to as `hadronic interactions'.

Decays of $K_L^0$ produce at most two charged particle tracks with the exception of a few rare decay modes. Of these the dominant decay, $K_L^0\to \pi^\pm e^\mp\nu e^+e^-$, contributes a branching ratio of $1.26\times10^{-5}$~\cite{bib:pdg2024}.
The tracks from $K_L^0$ decays are highly collimated given its typical boost factor $\gamma\gtrsim 20$.
Hence, the resulting DVs have a small opening angle of $\order{1^{\circ}}$, which makes them easy to separate from a potential LLP signal that is typically characterised by larger opening angles due to their larger mass and hence smaller boost.
Furthermore, only a small fraction of $K_L^0$ that escape the \atlas detector decay inside the \anubis active volume because of their large $\gamma$ factor.
The $n$ decays, given their long lifetime, contribute negligibly inside the \anubis active decay volume, leaving hadronic interactions as the primary mechanism for $n$ to contribute background to \anubis. 

Hadronic interactions are the dominant background mechanism from SM LLPs in \anubis.
They typically produce a handful of charged hadrons that are registered as a DV with a sizeable opening angle, see Section~\ref{sec:bkg_signatures}.
Hence, the signature of hadronic interactions can be similar to that from LLP decays.
One of the key advantages of \anubis is its {\em air-filled} active decay volume, this dramatically reduces the background from hadronic interactions, which are linearly proportional to the nuclear~density of material traversed by the SM LLP. 
More precisely, the air-filled active decay volume of \anubis reduces the likelihood of a hadronic interaction by a factor of $\order{10^4}$ relative to the tile hadronic calorimeter of \atlas.
This can be understood by comparing the nuclear interaction length for air of $\LIx{Air}=7.5\times10^{4}~\cm$ to that of the tile calorimeter $\LIx{Tile}=30.3~\cm$, steel $\LIx{Fe}=16.8~\cm$, and aluminium support structures $\LIx{Al}=39.7~\cm$~\cite{bib:pdg2024}.

%%%%%%%%%%%%%%%%%%%%%%%
\subsection{Background reduction using ATLAS}
\label{sec:bkg_reduction}
The potential background contribution from SM LLPs is substantially attenuated by $\order{10^{5}}$ by the \atlas detector acting as passive shielding.
The calorimeters alone account for $9.7~\LI$ at $\eta=0$~\cite{Aad:1129811}, resulting in an attenuation factor of $6.1\times10^{-5}$.
Also taking into account the support structures, the solenoid, and other material upstream of the calorimeters provides a total shielding equivalent to $11.2~\LI$ at $\eta=0$ and increasing to $13.1~\LI$ at $|\eta|=0.7$~\cite{Aad:1129811}. The flux of SM LLPs emanating from the interaction point is then reduced to $1.4\times10^{-5}$ and $2.0\times10^{-6}$ of its original value, respectively.

While the attenuation of the SM LLP flux emanating from the interaction point by the \atlas detector material is quite substantial, the surviving flux of $K_L^0$ and $n$ remains considerable at the HL-LHC given the total inelastic cross section of about 80~mb~\cite{bib:xsectotem,bib:xsecalfa,bib:xseclhcb}, which is dominated by QCD multijet processes.
This source of potential background can be further reduced using topological and kinematic selections on the prompt part of the $pp$ collision event registered by the \atlas detector, \ie using the \atlas detector as an {\em active veto} as outlined in the original \anubis proposal~\cite{bib:anubis_orig}.
In the first tier of the active veto, a selection on the missing transverse momentum reconstructed by the \atlas experiment of $\met>30~\GeV$ is applied. 
This requires that the SM LLPs that escape detection by the \atlas calorimeter,
and hence could initiate a hadronic interaction inside \anubis' active volume, 
be quite energetic, carrying a transverse momentum of $\pt\gtrsim30~\GeV$. 
Subsequent active veto selections exploit the fact that SM hadrons, including $K_L^0$ and $n$, tend to be produced in association with other hadrons that are reconstructed as jets. 
In particular, events with candidate LLP DVs in close angular proximity $\dR\equiv\sqrt{\Delta\eta^2+\Delta\phi^2}$ to energetic prompt jets with $\pt>15~\GeV$ reconstructed by the \atlas detector are vetoed if $\dR(\text{DV,jet})<0.5$.
In the same way, a similar veto is applied considering prompt charged particle tracks with $\pt>5~\GeV$ if $\dR(\text{DV,track})<0.5$.
The latter angular isolation requirement aims to eliminate background contributions from SM LLPs inside jets with few high-momentum particles.

The active veto using the information from the \atlas detector reduces the potential background contribution from SM LLPs by several orders of magnitude.
In addition to reducing the SM LLP flux from the interaction point, the angular isolation requirements on the DV are essential to remove any background contributions from non-prompt $K_L^0$ or $n$ production in hadronic showers that are initiated in the \atlas calorimeters, since these showers would be registered by the calorimeter and reconstructed as jets.

%%%%%%%%%%%%%%%%%%%%%%%
\subsection{Background predictions for ANUBIS}
\label{sec:bkg_predictions}

The dominant hadronic interaction backgrounds at \anubis are challenging to model using Monte Carlo (MC) simulations at the required precision level.
Hence, a data-driven approach is used to estimate the background contributions.

The background estimate follows the strategy of the original \anubis proposal~\cite{bib:anubis_orig} and is based on a search for LLPs with decay lengths of $\order{3~\metre}$ using ~36~\invfb of $pp$ collisions at $\sqrt s=13~\TeV$ recorded by the \atlas detector~\cite{ATLAS:2018tup}.
The search identifies LLP candidates through DVs reconstructed in the \atlas muon spectrometer (MS), which makes it very similar to \anubis in terms of detector technology, analysis strategy, targeted models, and $\sqrt s$.
In fact, this \atlas search inspired the active veto selection outlined in Section~\ref{sec:bkg_reduction}.
The background estimate for \anubis follows the DV+$\met$ analysis strategy in the barrel region, defined by $|\eta|<0.7$, which is most similar to \anubis in terms of kinematics, backgrounds, and instrumentation.
This \atlas search observed $N_{\rm bgr,ATLAS}=224$ events, which is consistent with the data-driven background estimate of $243\pm38~\stat\pm29~\syst$ obtained using uncorrelated sidebands, also known as the ABCD method~\cite{ATLAS:2018tup}.

The \anubis background estimate is obtained by scaling $N_{\rm bgr,ATLAS}$ to the conditions expected for \anubis:
\begin{eqnarray}
\label{eq:bg}
N_{\rm bgr, ANUBIS} 
&=& \frac{ \mathcal{L} }{ \mathcal{L'} }
\times \frac{ \LIx{MS} }{ \LIx{Air} }
\times \frac{\varepsilon}{\varepsilon'} 
\times \frac{ \leff }{ \leff' } \nonumber\\
&&\times f_{\rm bgr,MS}
\times N_{\rm bgr, ATLAS}\,,
\end{eqnarray}
where $\mathcal{L}$ is the integrated luminosity, $\eps$ denotes the reconstruction efficiency, $\Lambda_I$ represents the nuclear interaction length of a given region, $f_{\rm bgr,MS}$ is the fraction of \atlas backgrounds produced in the MS, and primed quantities refer to \atlas and the non-primed ones to \anubis.
Equation~\eqref{eq:bg} captures the differences in the active detector volume, $V$, through the effective path length $\leff\equiv\int_{\rm V} \dif\Omega \dif\ell$, \ie the integral over the solid angle, $\Omega$, and the path, $\ell$, where the path element, $\dif\ell$, points away from the interaction point, since the angular flux of SM LLPs $\frac{ \dif^2 N}{\dif\Omega dt}$ is constant as a function of distance in the absence of scattering.
This conservative definition of \leff ignores the decrease of the SM LLP flux with $\ell$ as it is depleted by hadronic interactions.

The effective path length for \atlas is calculated following Ref.~\cite{bib:anubis_orig} as:
\begin{eqnarray}
\leff'
&=& \int_{0}^{2\pi}\int_{0.92}^{2.22}\int_{R_\text{in=4.3\,m}}^{R_{\rm out=7\,m}} \dif\phi \dif\theta\sin\theta \dif\ell \nonumber\\
&=& 22.1~\metre\,,\label{eq:leff_atlas}
\end{eqnarray}
where the $\theta$ boundaries are defined by $\eta=\pm0.7$ and the path element d$\ell$ is conservatively bounded by the region between the innermost layer of the MS and the outermost radius before the DV reconstruction efficiency significantly drops~\cite{bib:dvalgo}. 

\begin{figure*}[tb]
\centering
\begin{subfigure}{0.45\textwidth}
\includegraphics[width=\linewidth]{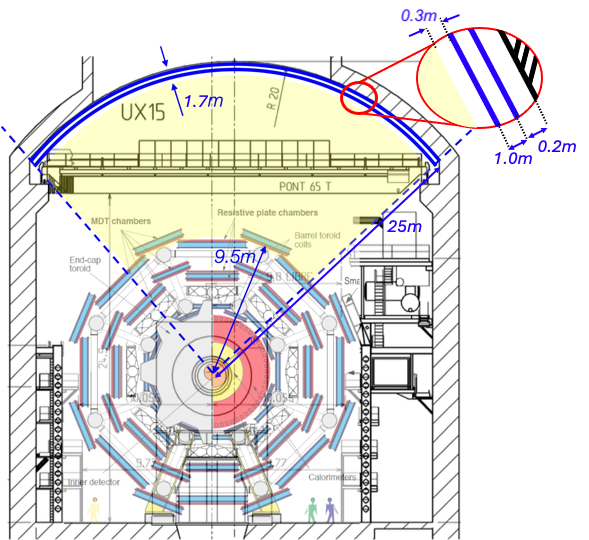}
\caption{}
\end{subfigure}\qquad
\begin{subfigure}{0.45\textwidth}
\includegraphics[width=\linewidth]{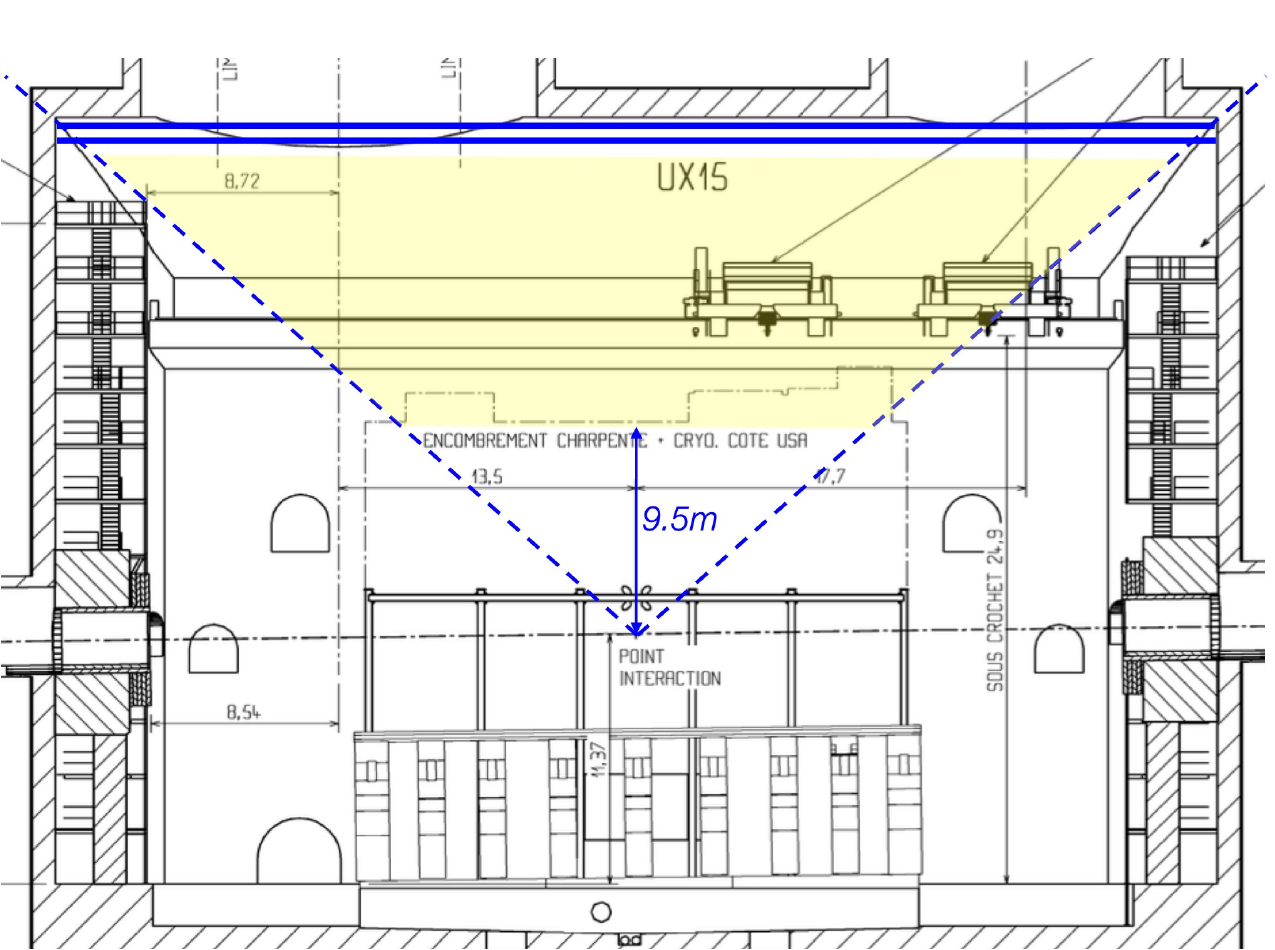}
\caption{}
\end{subfigure}
\caption{
\label{fig:ceilingDrawing}
The two layers of the tracking stations of the \anubis detector for the ceiling configuration are shown inside the UX15 cavern including the \atlas detector in (a)~the $(x,y)$ plane and (b)~the $(y,z)$ plane. 
The active volume is indicated as a shaded yellow region and the acceptance boundaries in $\eta$ and $\phi$ are highlighted as dashed blue lines. 
The drawings of the \atlas detector and the cavern courtesy of \atlas.
}
\end{figure*}

For \anubis in the ceiling configuration, the effective path length is similarly calculated as:
\begin{eqnarray}
\leff
&=& \int_{V} \frac 1{|\vec r|^2} \dif^3\vec r\nonumber\\
&=& 32.3~\metre\label{eq:leff_anubis}
\end{eqnarray}
Here, $V$ is bounded towards the beamline by the outer layer of the \atlas muon spectrometer~(9.5~m). 
The ceiling of the \atlas cavern dome defines the $V$ boundary away from the beamline.
The cavern dome has a curvature radius of 20~m, where the beamline is displaced by 1.7~m towards the centre of the LHC ring and by 3.5~m downwards relative to the centre of curvature.
The crossing points of the ceiling dome with the walls of the \atlas cavern define the radial $V$ boundary through direct lines connecting them to the interaction point.
A drawing of the \anubis detector in the ceiling configuration inside the UX15 cavern including the \atlas detector is shown in Figure~\ref{fig:ceilingDrawing}, where the active volume considered for the background estimate is indicated as a shaded region and the acceptance boundaries in $\eta$ and $\phi$ are highlighted as dashed lines.

Background events from hadronic interactions in the \atlas search include DVs produced in both the outer layers of the \atlas calorimeter and the \atlas MS. 
The reconstruction efficiency of DVs from hadronic interactions in the calorimeter displays a strong variation as a function of $\ell$. 
This is due to a combination of decay products being stopped within the calorimeter before reaching the MS and observed deposits in the calorimeter causing the events to fail isolation requirements.
By contrast, the reconstruction efficiency of DVs from hadronic interactions in the MS is relatively stable as a function of $\ell$ until it drops around 7~m~\cite{bib:dvalgo}.
Hence, only the events with DVs that are consistent with being produced in the MS are used for the \anubis background prediction. 
Their fraction relative to the total number of \atlas background events is captured by $f_{\rm bgr,MS}=0.65$, calculated by measuring the relative longitudinal density of vertices seen in \atlas within the calorimeter and MS and correcting for vertex migration due the finite precision of the DV reconstruction algorithm.

The average nuclear interaction length within the \atlas MS, \LIx{MS}, is calculated using the average density of aluminium and iron in the MS, measured from the \atlas geometry simulation~\cite{ATLAS:2020ell}. 
Conservatively, any materials within the MS other than aluminium and iron are neglected, resulting in \LIx{MS} = 385 \cm.

With the estimates of the effective path length from Eqs.~\eqref{eq:leff_atlas} and \eqref{eq:leff_anubis}, using a projected HL-LHC integrated luminosity of 3~\invab, and conservatively assuming that, due to their uniform coverage, the \anubis detectors will be two times more efficient than the current \atlas muon system in the context of reconstructing background DVs, Eq.~\eqref{eq:bg} yields:
%\begin{widetext}
\begin{eqnarray}
N_{\rm bgr, ANUBIS\,ceil.} 
&=& 0.86%3
\times \frac{ \int_{V=\rm ceil.} \dif\Omega \dif\ell }{ \int_{V'=\rm ATLAS} \dif\Omega \dif\ell } \nonumber\\
&&
\times f_{\rm bgr,MS}
\times N_{\rm bgr, ATLAS} \nonumber \\
&=& 0.56 
\times \frac{ 32.3\,\metre }{ 22.1\,\metre } 
\times N_{\rm bgr, ATLAS} \nonumber\\
&=& 182.4 \pm 12.2\,.
\label{eq:bgnum}
\end{eqnarray}
%\end{widetext}
As a cross-check of the above technique, an alternative background prediction was made by rescaling the interactions within the \atlas calorimeter instead of those in the MS. 
In this estimate, the dependence of the DV reconstruction efficiency on the distance from the beamline was explicitly taken into account. 
Using the efficiencies from the \atlas search for a range of signals,
along with the relative volume of calorimeter and its average material density, the calorimeter-based estimates predict backgrounds ranging from 
19 to 107 events.
As the MS-based estimate produces the more conservative prediction and has a lower uncertainty due to not needing to estimate the differential vertex reconstruction efficiency, it is used as the primary background prediction.
Given the conservative assumptions and taking the MS-based background estimate rather than the calorimeter-based one, 
these projections will be referred to as `conservative' in the following.

Recently, \atlas has published a new search for LLPs with decay lengths of \order{3~\metre} using 140~\invfb of $pp$ collision data at $\sqrt s=13~\TeV$~\cite{ATLAS:2025pak}.
This new analysis is conceptually similar to the results using 36~\invfb of data from Ref.~\cite{ATLAS:2018tup}, but achieves an even higher background rejection rate through machine learning techniques, and gives the background contribution in the barrel region as  $241 \pm 18~\stat \pm 6~\syst$, which is consistent with the observation of $N_{\rm bgr, ATLAS, 140\,\invfb} = 245$.
Using this background estimate in Eq.~\eqref{eq:bgnum} and accounting for the different ratio of integrated luminosities between the \atlas analyses gives
\begin{equation}
N_{\rm bgr,\,ANUBIS\,ceil.}^{\rm from 140~\invfb} 
= 51.3 \pm 3.2\,.
\label{eq:bgnumnew}
\end{equation}

%While the projection from Eq.~\ref{eq:bgnumnew} underlines the potential sensitivity of ANUBIS to LLP signals that could be achieved using multivariate machine learning techniques, it is not further used in this paper and instead the results from Eq.~\eqref{eq:bgnum} are considered for the `conservative' projections.
While the estimate from Eq.~\eqref{eq:bgnumnew} underlines the potential sensitivity of \anubis to LLP signals that could be   achieved using multivariate machine learning techniques and hence supports `nominal' projections, the results from Eq.~\eqref{eq:bgnum} are considered for the `conservative' projections.
 
For \anubis in the shaft configuration, which corresponds to the `shaft+cone' scenario in the original \anubis proposal~\cite{bib:anubis_orig}, the effective path length was calculated as $\leff=11.3$~m. 
This results in 
\begin{equation}
N_{\rm bgr,\,ANUBIS~shaft+cone} 
=63.7 \pm 4.3\,.
\label{eq:bgnumshaft}
\end{equation}

The actual number of background events in the HL-LHC dataset is likely to be lower than indicated in Eqs.~\eqref{eq:bgnum} and~\eqref{eq:bgnumnew}. 
To provide a bound on the best-case \anubis sensitivity, a `background-free' projection is also provided in  Section~\ref{sec:SensitivityResults}.
Hence, the actual sensitivity of \anubis is bracketed by the `background-free' projections and the `conservative' projections based on Eq.~\eqref{eq:bgnum}.
Note that searches requiring the presence of two LLPs in the combined active volume of \anubis and \atlas can safely be considered background-free. 
The same is true for scenarios where LLPs are produced in association with prompt energetic SM particles like \eg a $W$ or $Z$ boson that are required to be registered by \atlas.

%%%%%%%%%%%%%%%%%%%%%%%
\subsection{Signature of hadronic interactions in ANUBIS}
\label{sec:bkg_signatures}

As outlined in Section \ref{sec:bkg_sources}, SM LLPs represent the dominant potential source of background for \anubis: both $K_L^0$ and $n$ can contribute through hadronic interactions, while decays are only relevant for the former.
Despite low background levels presented in Section~\ref{sec:bkg_predictions}, it is prudent to study in detail the topology of contributions from SM LLPs to understand what signature they will present to the \anubis detector.

As reported in Ref~\cite{satterthwaite_2022_292b9-eck73}, a sample of minimum-bias $pp$ collision events at $\sqrt s=14~\TeV$ is generated using \pythia[8]~\cite{Sjostrand:2014zea} using the MSTW2008 parton distribution functions~\cite{bib:mstw} with the A2 tune~\cite{bib:a2}.
Out of $10^{10}$ simulated collisions, approximately $2.3 \times 10^7$ $n$ and $1.7 \times 10^7$ $K_L^0$ particles are produced.
In the next step, a separate simulation is carried out using \geant~\cite{GEANT4:2002zbu} to model $n$ and $K_L^0$ interactions by firing them into a coaxial air-filled cylinder of 100~m in length and 20~m in radius.
In this simulation SM LLPs are generated following the three-momentum distributions in the minimum bias event sample generated in the first step.

The \geant simulations show that in the majority of events the SM LLP particles pass through the air-filled cylinder, which is expected given $\LIx{Air}$.
In the remainder of the events, the SM LLP particles undergo a hadronic interaction, with a small fraction of $K_L^0$ decaying.
Only events with at least two charged particles in the final state are retained for further analysis, since no DV can be reconstructed otherwise. 
The resulting sample of events will be referred to as the `interaction sample'.

The results of the \geant and \pythia[8] simulations are combined for further analysis.
For each $n$ or $K_L^0$ in the minimum bias sample, a corresponding particle with a three-momentum within 50 MeV of the minimum bias particle is randomly selected from the interaction sample.
The resulting event is then weighted according to the probability of generating the required multiplicity of charged final-state particles, based on the fraction of events in the interaction sample within the 50~MeV momentum window that met this condition.
If no kinematic match is found, the three-momentum acceptance threshold is increased in 50~MeV increments until a match is found. 

The trajectories of final-state charged particles produced in the step above are subject to an angular transformation that aligns the three-momenta of the SM LLP in the interaction sample to the corresponding matched SM LLP in the minimum bias sample.
Hence, the \pythia[8] simulation of minimum bias events contributes the type of the SM LLP as well as its trajectory and event-level properties like \met, while the \geant simulation of the interaction sample supplies details on the distance travelled by the SM LLP before interaction with air or decay, and the spectrum of particles produced.
The resulting events were subjected to the event-level selection criteria from Section~\ref{sec:bkg_reduction}: $\met > 30~\GeV$ and $\Delta R > 0.5$ between the DV and the closest reconstructed jet or charged particle. 

The charged particle multiplicity in the selected background event sample originating from hadronic interactions peaks at four and shows an exponentially falling tail extending up to 20.
These numbers include low-energetic secondary $e^\pm$ from converted Bremsstrahlung photons, $\pi^0\to\gamma\gamma$ decays, etc.
The angular distribution between the trajectories of the charged particles and the original trajectory of a SM LLP steeply increases up to an angle of about 0.2 rad and falls exponentially after that.
For the $K_L^0$ decays, the charged particle distribution is dominated by a peak at two, and has an angular distribution peaking at 0.02 and extending up to 0.05. 
Hence, $K_L^0$ decays can be easily discriminated against using their signature of two very collimated charged particle tracks. 
However, the topology of potential background events from hadronic interactions can mimic the topology of the signal in principle.
This underlines the pivotal advantage of the air-filled active decay volume of \anubis, which allows for maximal reduction of the contribution from hadronic interaction.  

To study the potential impact of hadronic interactions in the context of the geometry of the \anubis detector, the trajectories of charged particle tracks are propagated to determine if they would intersect with the \anubis tracking stations.
At least two charged particles 
are required to intersect with the tracking stations, since at least two reasonably energetic tracks are required to produce a reconstructable displaced vertex.
This approach assumes the detector to be 100\% efficient for charged particles.
This approximation is made since a single RPC detector features a detection efficiency of $>$98\%, which translates into $\sim$99.9\% for a tracking layer that comprises a triplet of RPC detectors with a two-out-of-three hit logic.

\enlargethispage{5mm}

\begin{figure*}[tb]
    \centering
    \begin{subfigure}[b]{0.9\textwidth}
        \centering
        \includegraphics[width=0.85\linewidth]{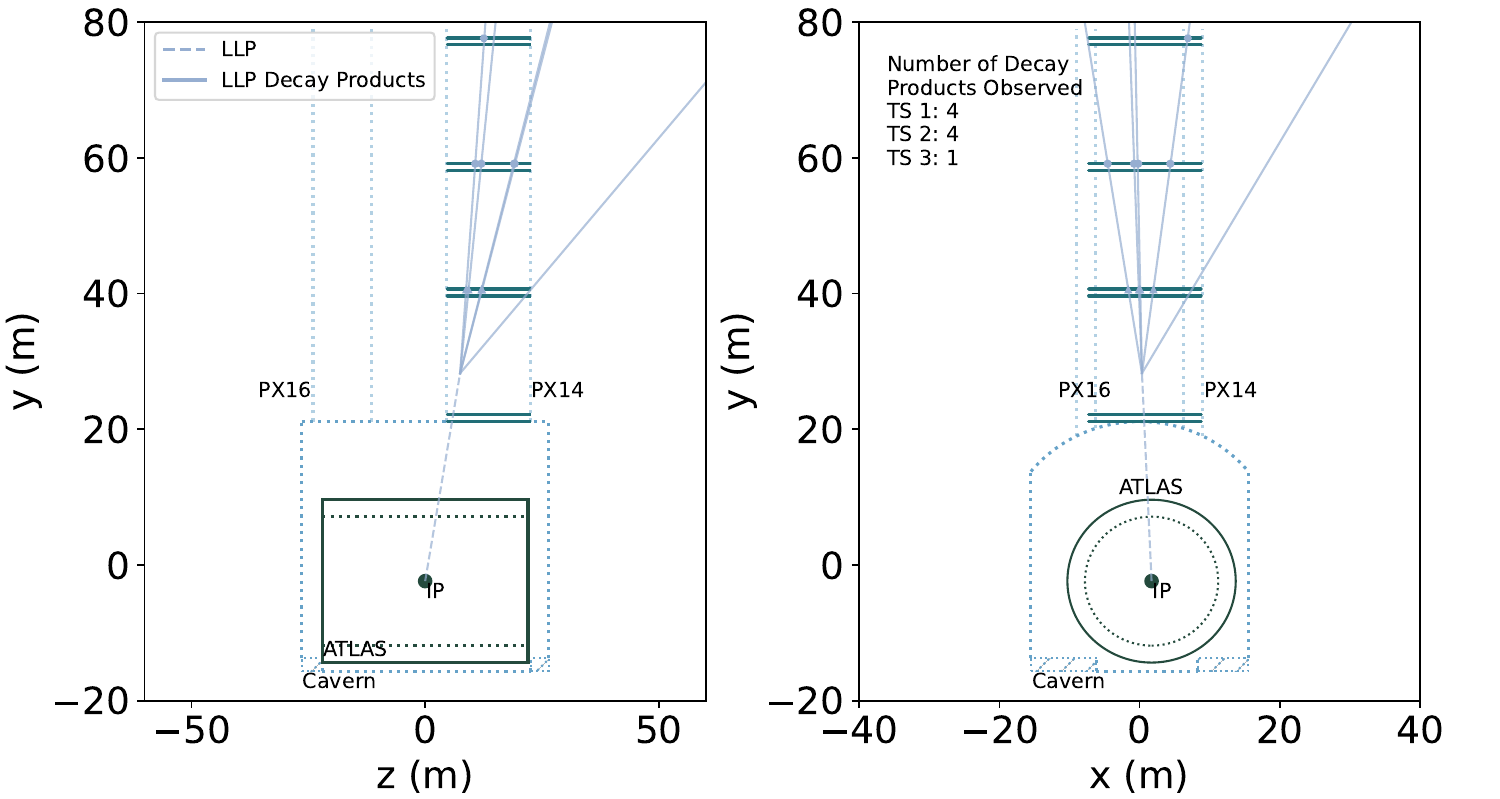}
        \includegraphics[width=0.95\linewidth]{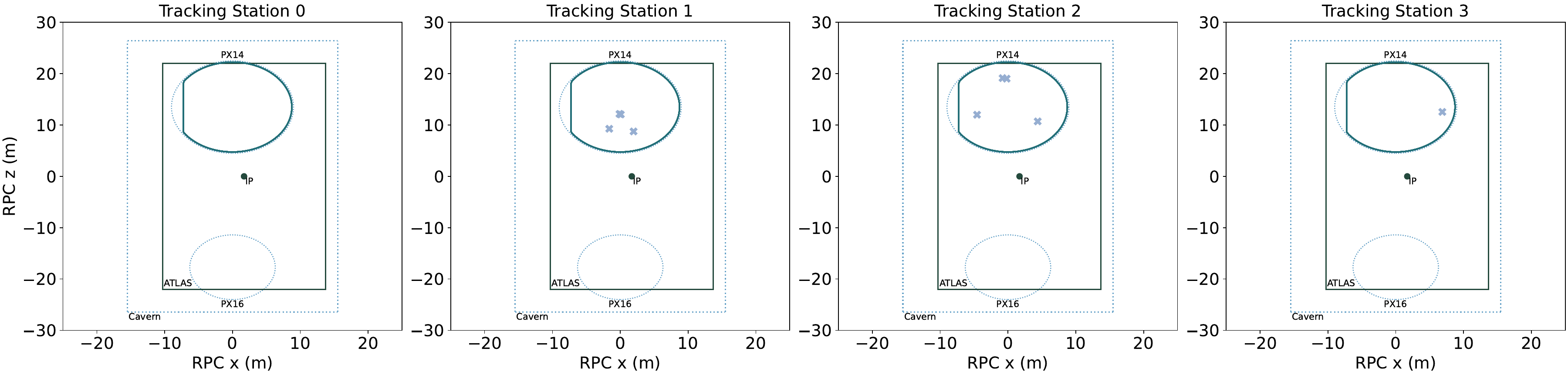}
\end{subfigure}
\caption{
\label{fig:CeilingFirstShaftConfigBackgrnd}
Display of a representative background candidate event featuring a hadronic interaction induced by a $n$ inside the \atlas Cavern. 
Dashed lines represent the LLP trajectory, while solid lines show the trajectories of the ensuing jet of 4 charged final-state particles. 
Cross-sections of the \atlas cavern and service shafts in the $zy$- (top left) and $xy$- (top right) planes, with dotted lines representing \atlas' vertexing limit. 
The $xz$ intersection points of the final-state, charged particles with \anubis' tracking stations inside the shaft (bottom) are shown as crosses.
}
\end{figure*}

A representative background candidate event involving a hadronic interaction of a SM LLP, in this case a $n$, is displayed in Figure~\ref{fig:CeilingFirstShaftConfigBackgrnd}.
In this event, five charged final-state particle tracks are produced inside the shaft between the first and second TS. 
The second and third TS register hits from four tracks, while the fourth TS observes hits from two.
No track hits are registered in the ceiling tracking station in this particular event.
The opening angle of the charged particles emanating from the DV in this example event is fairly small, as expected for hadronic interactions.

\begin{figure}[tb]
\centering
\begin{subfigure}[b]{\columnwidth}
\includegraphics[width=0.95\linewidth]{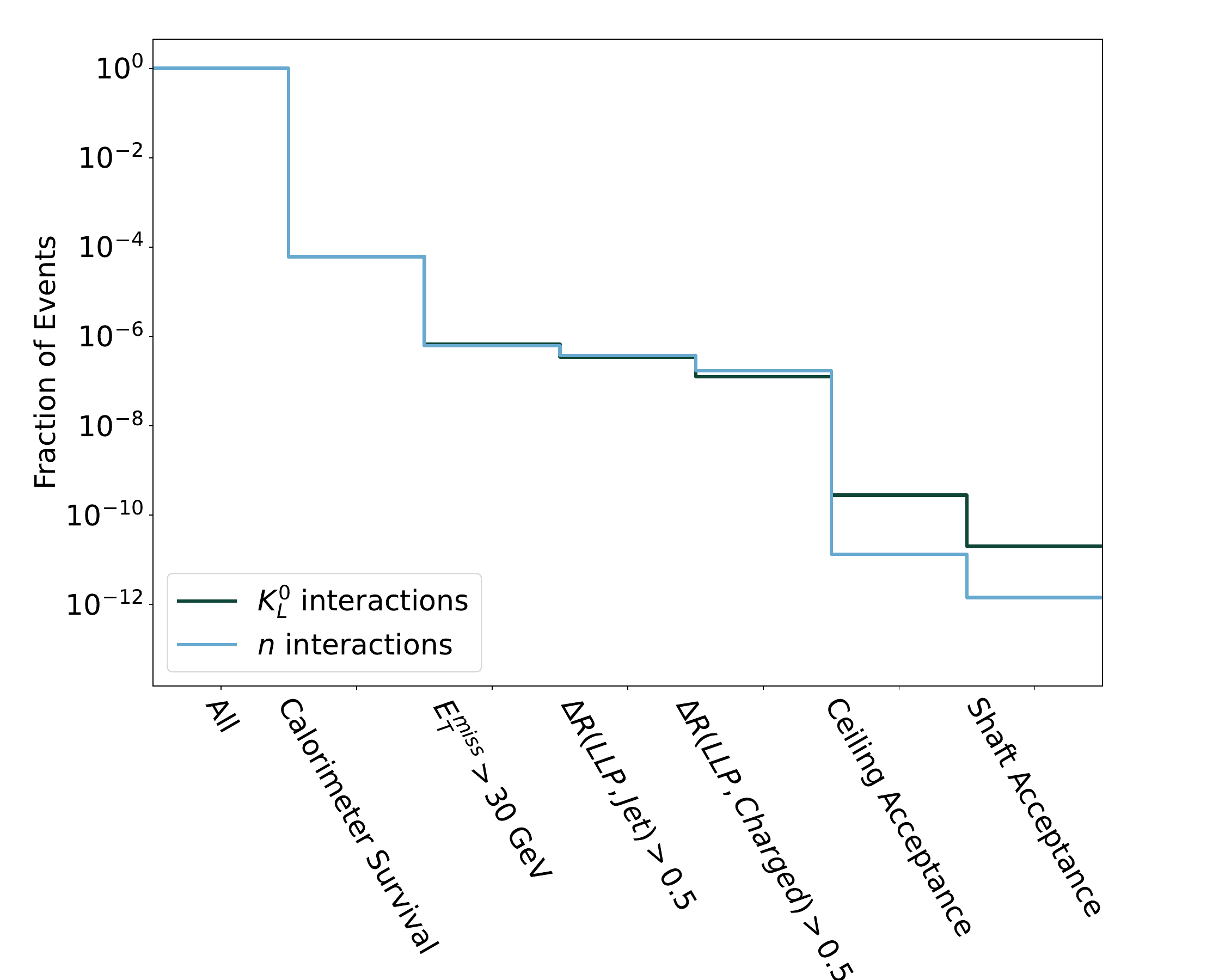}  
\end{subfigure}
\caption{
The impact of successive selection requirements and background events from $K_{L}$ and $n$ interactions within the active volume of the \anubis detector. 
Note that the final two bins are non-cumulative: events passing the final event-level selection are considered independently for the `ceiling' and `shaft' scenarios respectively. 
} 
\label{fig:CutFlowSignalBackg}
\end{figure}

The impact of topological selections from Section~\ref{sec:bkg_reduction} on the simulated background events is studied in 
Figure~\ref{fig:CutFlowSignalBackg}.

As expected, the shielding effect of the \atlas calorimeters, which is conservatively\footnote{This is because the support structures are not considered here, if they are then the calorimeters should be taken as 11.2~\LI.} taken as 9.7~\LI~\cite{Aad:1129811} , has the most prominent impact on the background, reducing it by $6\times10^{-5}$. 
The second most prominent selection is the $\met>30~\GeV$ requirement, which reduces the background by another two orders of magnitude.
The angular isolation requirements then reduce the background by another order of magnitude.
The requirement for the SM LLP interaction to occur within the active volume of \anubis further reduces the background by three to four orders of magnitude, depending on the detector configuration.

Normalising to the product of the inelastic proton-proton cross-section of 78.1~mb and ${\mathcal{L}= 3 \text{ ab}^{-1}}$~\cite{Apollinari:2015bam}, the above predictions are larger than the conservative data-driven estimate in Section~\ref{sec:bkg_predictions}.
This disagreement is attributed to a combination of factors like the simulations not considering selections that would be performed in an actual analysis, \eg vertex reconstruction or track or hit multiplicities, as well as potential modelling deficiencies of rare processes like hadronic interactions with air in \pythia[8]{}+\geant simulations.

To study these factors, a prototype demonstrator called \proanubis consisting of one \anubis TS module was installed in 2023 close to the ceiling of the \atlas cavern and has so far recorded 157.8~\invfb of $pp$ collision data~\cite{Shah:2024fpl}.
One of the main physics goals of \proanubis is the direct measurement of the expected background flux in a background-enriched region, which then can be used to refine the vertexing algorithm and other DV-related selections, background simulations, and further optimisations.

%%%%%%%%%%%%%%%%%%%%%%%%%%%%%%%%%%%%%%%%%%%%%%
\section{Sensitivity Studies} 
\label{sec:SensitivityStudies}
\subsection{Signal Model}
\label{sec:SensitivitySignalmodel}
The sensitivity of \anubis to detect BSM neutral LLPs is evaluated through simulations of their production and decays.
The primary benchmark model considered involves the Higgs boson decaying into a pair of neutral, scalar LLPs, $h\rightarrow ss$, where $s$ denotes a new neutral scalar particle that may be long-lived.
Throughout this article, $s$ and $m_s$ denote this scalar state and its mass, respectively.

For $m_s>10~\GeV$, each scalar is assumed to decay predominantly into a pair of $b$-quarks, corresponding to Benchmark Case 5 (BC5) from Ref.~\cite{Beacham:2019nyx}.
These quarks then hadronise, producing charged particles that can be detected by the tracking layers of \anubis.
The detection of these final-state jets provides the key signature of the signal.

The choice of exotic Higgs decays into neutral LLPs as the benchmark is motivated by their prevalence in BSM scenarios, including neutral naturalness~\cite{Giudice:1998bp,Burdman:2006tz,Argyropoulos:2021sav} and supersymmetric models, where LLPs often arise.
Furthermore, this model features a massive mediator at the electroweak scale -- the Higgs boson, thereby highlighting the unique sensitivity of transverse detectors like \anubis and their complementarity to forward detectors.
This benchmark model is endorsed by the Physics Beyond Colliders initiative for comparisons between various experiments~\cite{Beacham:2019nyx} features in its ESPPU 2026 submission~\cite{PBC:2025sny}.
Signatures with displaced hadronic jets similar to BC5 presented here arise in a broad class of BSM scenarios~\cite{Alimena:2019zri}.
%The analysis is consistent with existing searches conducted by the ATLAS and CMS experiments~\cite{Lee:2018pag, ATLAS:2018tup, CMS:2014hka}, allowing for meaningful comparisons and extrapolations of the ANUBIS performance relative to the current collider detectors.

\begin{figure}
    \centering
    \includegraphics[width=0.65\linewidth]{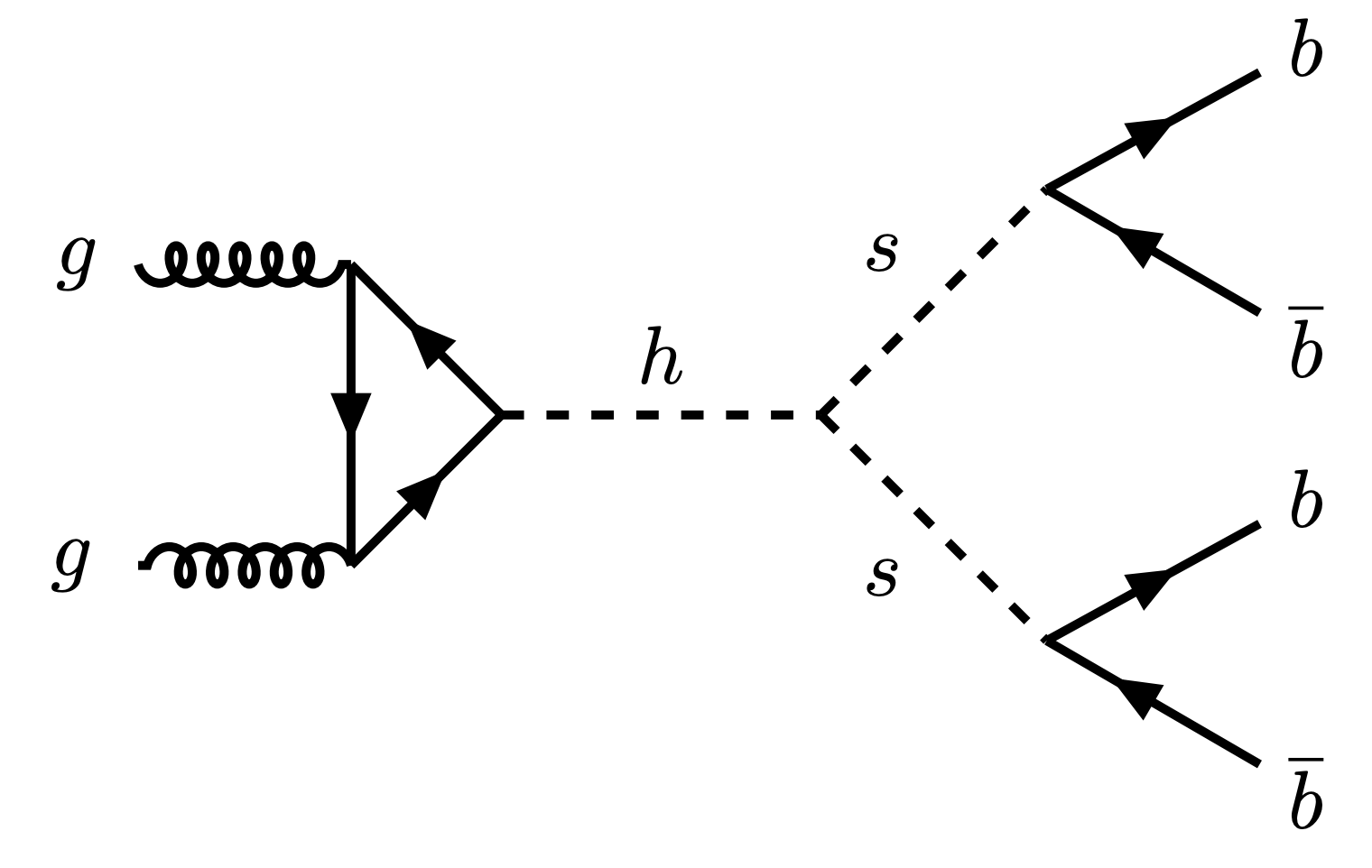} 
    \caption{Feynman diagram for the dominant scalar LLP production mechanism considered in this analysis, which is via a Higgs boson produced with gluon-gluon fusion.
    } 
    \label{fig:SignalModel}
\end{figure}

This analysis considers the dominant Higgs boson production mechanism: gluon-gluon fusion (ggF), which accounts for 88\% of Higgs production, with a cross-section of 54.67~\pb at $\sqrt{s} = 14$ TeV~\cite{bib:pdg2024}.
The corresponding Feynman diagram is shown in Figure~\ref{fig:SignalModel}.
%Decays into $b\bar{b}$ pairs are considered due to the expectation that the LLP would exhibit Higgs-like Yukawa couplings, favouring decays into heavy fermions such as $b$-quarks.

\subsection{Signal sample simulation} 
\label{sec:SenSimulations}
The simulated signal samples were produced using the Hidden Abelian Higgs model (HAHM) described in Refs.\cite{Strassler:2006im,Han:2007ae,Curtin:2013fra,Curtin:2014cca} to generate a ggF-produced Higgs that is forced to decay into a pair of electrically neutral scalar ($s$) LLPs with masses of $m_s=10, 20, 30, 40, 50$, and 60~GeV. These were simulated using MG5\_aMC@NLO 3.5.4~\cite{Alwall:2014hca} with the NN23LO1 PDF set~\cite{Ball:2013hta}. The produced scalar LLPs are forced to decay into a pair of b-quarks with a branching ratio $\br(s\rightarrow b \bar b)=100\%$ via Madspin, with the subsequent hadronisation and parton shower simulated with 
%version 8.311 of 
the \pythia[8] programme. % , using X PDFs and 1 tune (the defaults).   
In total, approximately $4\times10^6$ events were simulated for each mass point. The HAHM also includes an additional dark photon, $Z_d$~\cite{Curtin:2014cca}, which is decoupled here by setting its mass to an arbitrarily high value ($\gg m_h$) and its coupling to the higgs boson to be zero.

\subsection{Event Selection} 
\label{sec:SenEventSelec}
\begin{figure}[htpb]
\centering
    %\begin{subfigure}[b]{\textwidth}
    \includegraphics[width=0.9\linewidth]{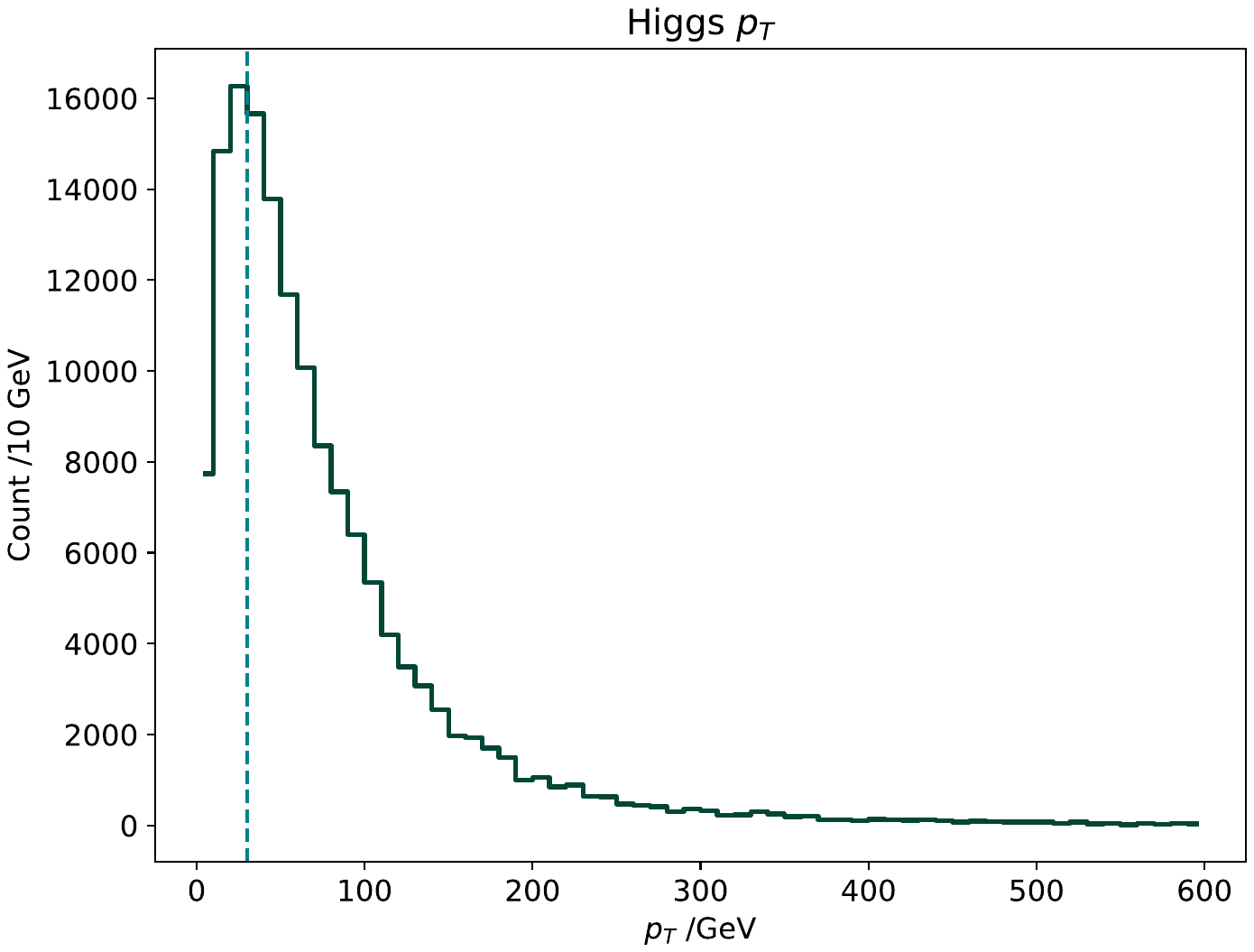} 
    %\includegraphics[width=0.5\linewidth]{HiggsPTdistribution_VBF.pdf}  
%\end{subfigure}
\caption{
\label{fig:HiggsPtDistributions}
Expected \pt distribution of the Higgs boson simulated for the ggF 
%(left) or VBF (right) 
production mode. The $\met>30~\GeV$ cut is indicated by the dashed vertical line.
}
\end{figure}

To identify candidate events in the signal samples consistent with neutral BSM LLP decays the selections from Section~\ref{sec:bkg_reduction} are applied: candidate LLP decays must be isolated by $\Delta R=0.5$ from any nearby jets with $\pt>15~\GeV$ and any energetic charged particles with $\pt>5~\GeV$; and $\met>30~\GeV$ is required in the event.
The \met selection is expected to have a moderate impact on the signal efficiency, as indicated by the \pt distribution of the Higgs boson in Figure~\ref{fig:HiggsPtDistributions}, leaving space for further tightening of the $\met>30~\GeV$ criterion. Additionally, to better represent the response on data, a geometric acceptance cut requires the LLP decay to occur within the \anubis fiducial volume; and that the momentum vectors of any charged decay products of the LLP intersect with at least two RPC stations. 

\begin{figure}[htpb]
    \centering
    \includegraphics[width=0.9\linewidth]{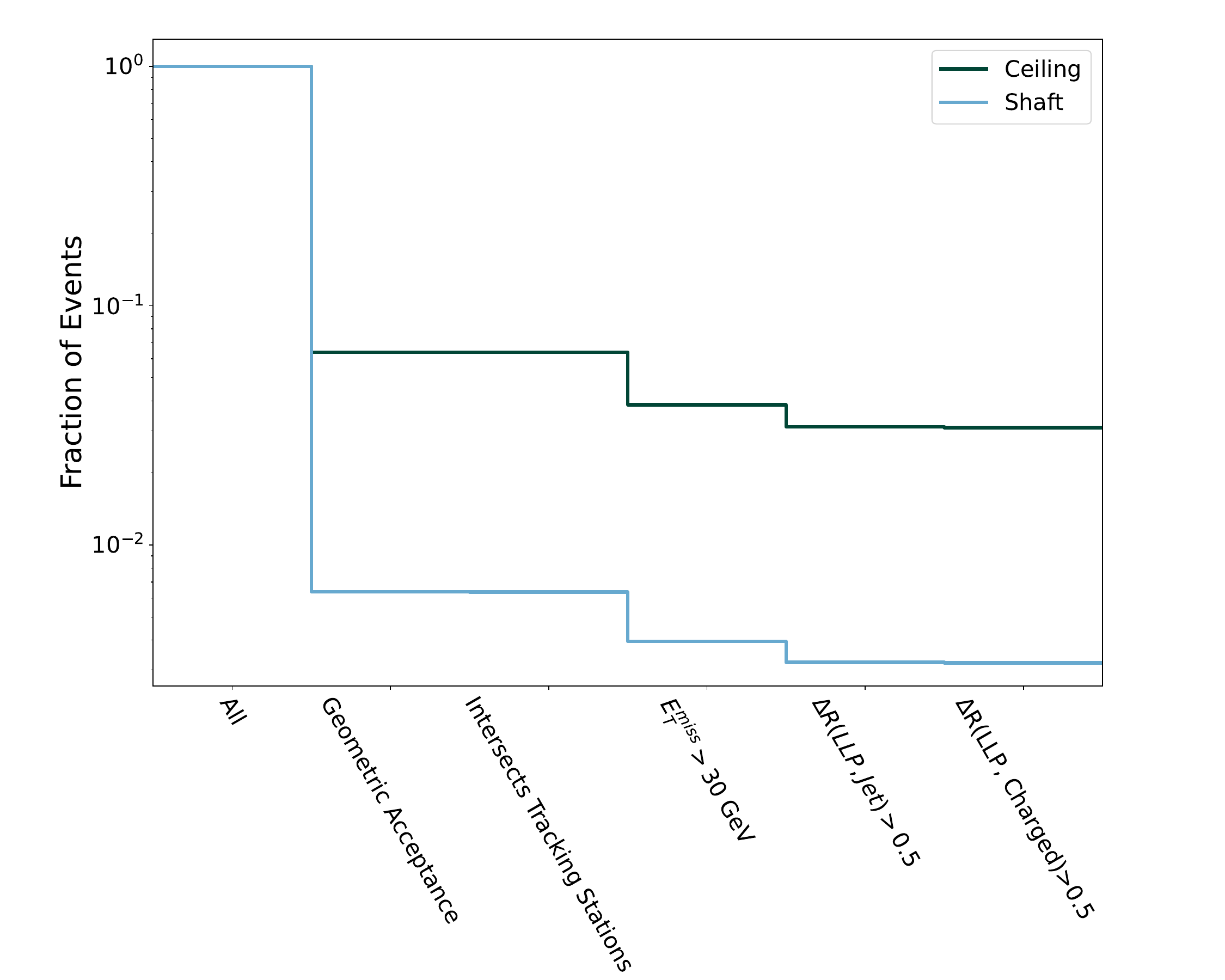}
    \caption{The impact of successive selection requirements on the signal sample with $m_s=15$ GeV and $c\tau=3$ m for the ceiling and shaft \anubis configurations. The ordering of these cuts differs from the background case, where the geometric acceptance was performed last for the ceiling and shaft separately.}
    \label{fig:CutFlowSignal}
\end{figure}

Figure~\ref{fig:CutFlowSignal} shows the impact of these selections on a signal sample with $m_s=15~\GeV$ and $c\tau=3~\metre$. Following the application of these selection cuts, the fraction of signal events that are retained is approximately 0.03 in the ceiling case and 0.003 in the shaft case, while the fraction of SM LLP backgrounds that are expected to survive is $\mathcal{O}(10^{-12})$ and $\mathcal{O}(10^{-11})$  for the respective geometries. This highlights the balance between signal purity and background suppression from the selection. 

This selection considers a single displaced vertex (1DV) within the \anubis detector, however there are two scalars in the final state. Therefore, an additional constraint that there must be two displaced vertices (2DVs) in an event can be introduced, where at least one of these vertices is within the \anubis fiducial volume and the other is in \atlas or \anubis. This requirement reduces the background level to effectively zero, since if the fraction of remaining background events is $\approx10^{-12}$ for the 1DV selection, which translates to $\approx10^{-24}$ for the 2DVs case.

\subsection{Detector geometry simulations} 
\label{sec:SenGeometricModel}
The simulation of the \anubis detector geometry and the corresponding acceptance for signal events is carried out following the same procedure as for background events described in Section~\ref{sec:bkg_signatures}.
%that are needed to determine the number of LLP-producing events that pass the selection criteria and generate final-state charged decay products intersecting the ANUBIS TSs are evaluated.
%Charged particles are conservatively required to have a minimum three-momentum of 100 MeV to ensure that they can be detected by both the inner and outer layer of a TS, which is necessary to reconstruct a track.
%
For an event to be observed, the LLP must produce at least two sufficiently energetic charged particles that intersect at least one of the \anubis TSs.
In addition, final-state charged particles that traverse dense materials such as concrete before reaching a TS are excluded from the analysis, as they would likely be absorbed.
%This requirement ensures that sufficient information is available to reconstruct the decay vertex of the parent LLP.
%The analysis does not account for the probability that energetic charged particles fail to be detected by the ANUBIS TSs given the high RPC detector efficiency of close to 100\% for triplet of detectors featuring a detection efficiency of  $>$98\% each combined following a two-out-of-three hit logic.
%
%For simplicity, the LLPs are assumed to be produced precisely at the interaction point, and the final-state particles from their decays are considered to emerge directly from the LLP decay vertex.
%These assumptions have a negligible impact on the results, which is expected given the large dimensions of ANUBIS. 
%are reasonable given the short lifetimes of both the Higgs boson (the progenitor of the LLP) and the LLP's decay products, making their effect on the analysis negligible.
The decay positions of simulated LLPs are reweighted according to 
\begin{equation} 
P(x) = \frac{1}{\beta\gamma c \tau} e^{-x/(\beta\gamma c \tau)}~\text{\cite{bib:pdg2024}}.
\label{Eq:ExpoDecay}
\end{equation} 
%where $c\tau$ ranges from $10^{-3}$ to $10^8~\metre$.

\begin{figure*}[htpb]
\centering
\begin{subfigure}[b]{0.99\textwidth}
\centering
\includegraphics[width=0.85\linewidth]{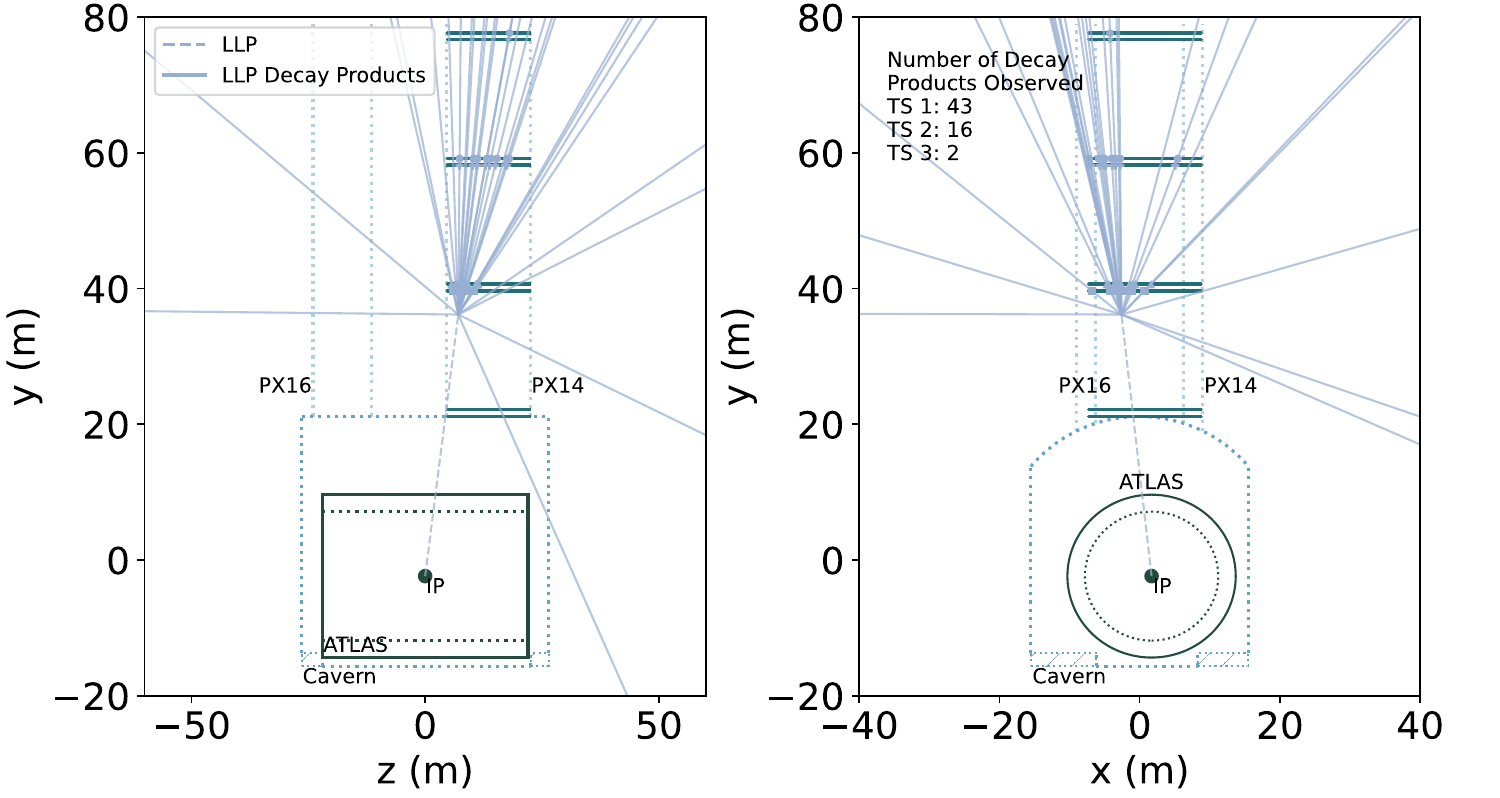}
\includegraphics[width=0.95\linewidth]{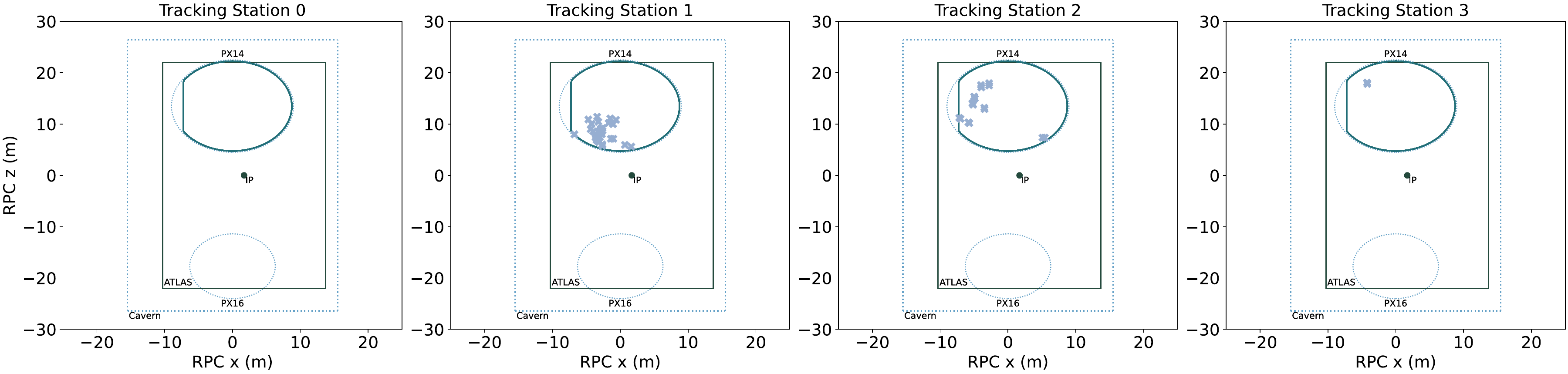}
\end{subfigure}
\caption{
\label{fig:ShaftConfigTS1}
Display of a representative signal event where the decay of a scalar LLP with $m_s=15~\GeV$ and $c\tau=30$ m is observed by \anubis in the shaft configuration after the full selection procedure. This event would not be observed in the ceiling configuration.
Dashed lines represent the LLP trajectory, while solid lines show the trajectories of the ensuing charged final-state particles produced in the LLP decay. 
Cross-sections of the \atlas cavern and service shafts in the $zy$- (top left) and $xy$- (top right) planes are also shown, with dotted lines representing \atlas' vertexing limit. 
The intersection points of the final-state, charged particles with \anubis' Tracking Stations in $xz$ projection (bottom) are shown, where each represents two triplets of RPCs separated by $\sim1$ m.
The intersection points are represented as crosses.
%Dashed lines represent the LLP trajectory, while solid lines show the trajectories of the ensuing jet of 14 charged final-state particles. 
%Cross-sections of the ATLAS cavern and service shafts in the $zy$- (top left) and $xy$- (top right) planes, with dotted lines representing ATLAS' vertexing limit. 
%The intersection points of the final-state, charged particles with ANUBIS' tracking stations are shown for the ceiling using local curved coordinates (bottom left) and for the tracking stations inside the shaft (other bottom row panels).
%The intersection points are represented as dark (light) crosses if they are within (outside) of the active area of a tracking station.
} 
\end{figure*}

\begin{figure*}[htbp]
\centering
\begin{subfigure}[b]{0.99\textwidth}
\centering
\includegraphics[width=0.65\linewidth]{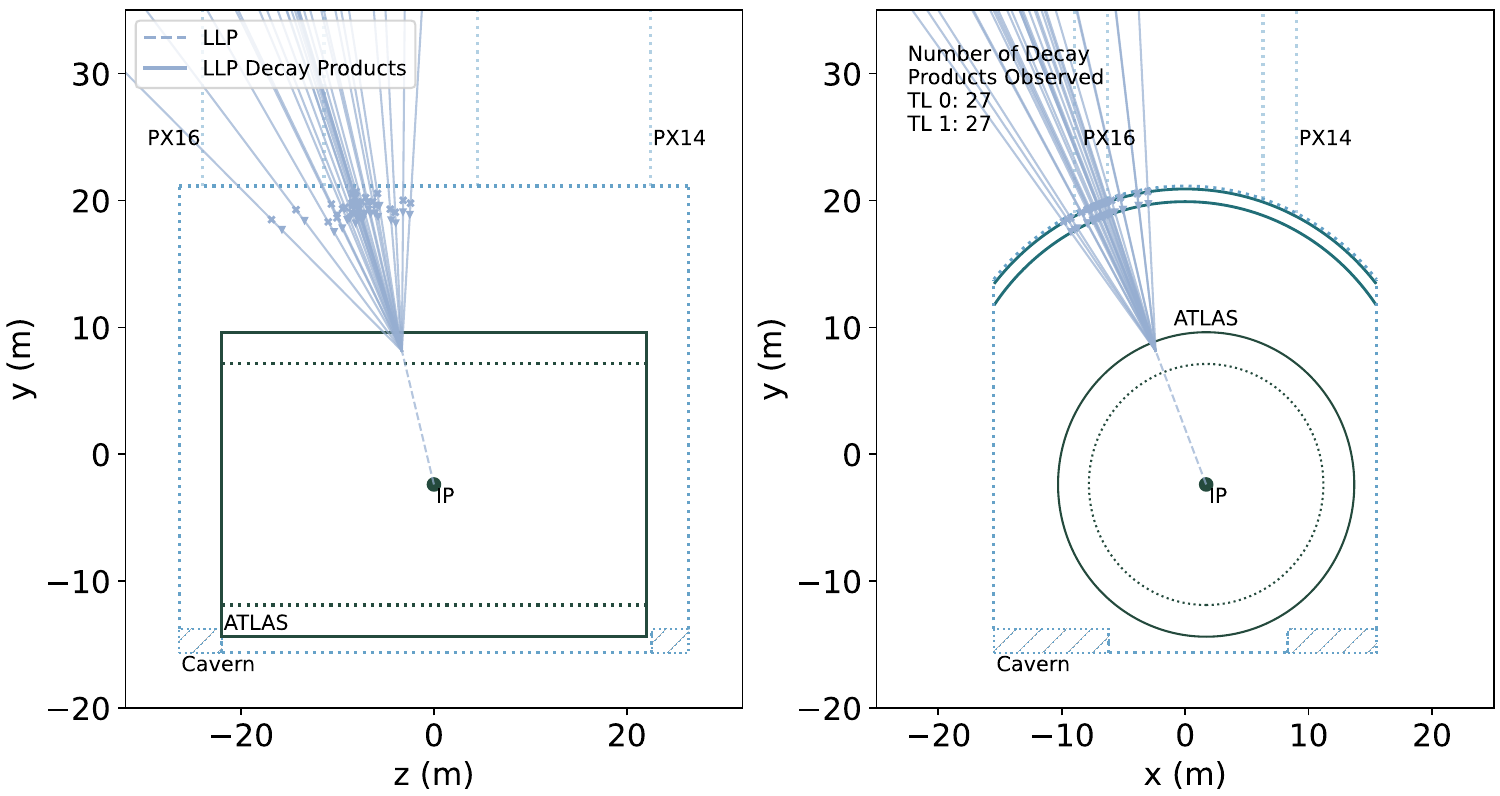}
\includegraphics[width=0.65\linewidth]{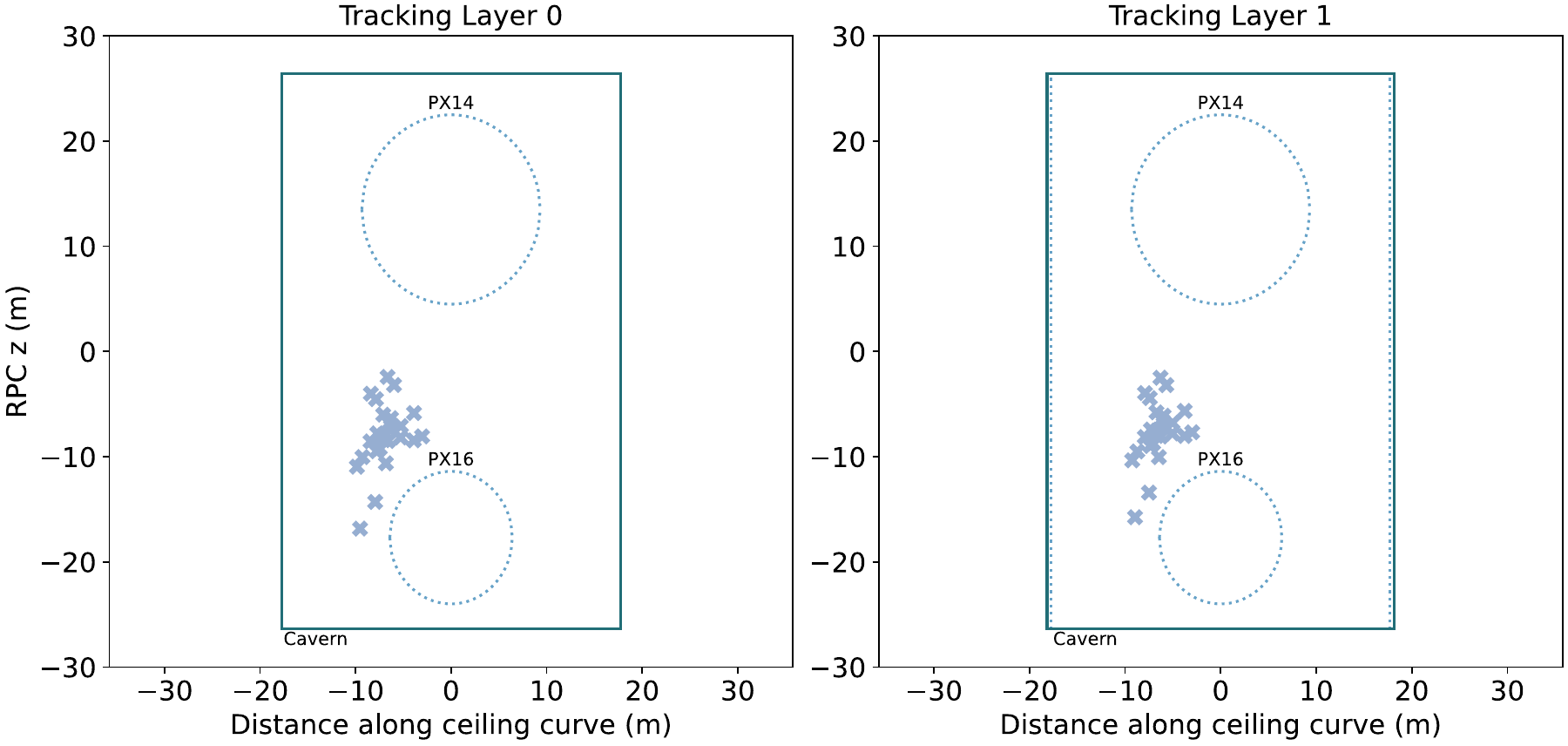}
\end{subfigure}
\caption{
\label{fig:CeilingConfig}
Display of a representative signal event where the decay of a scalar LLP with $m_s=15~\GeV$ and $c\tau=3$ m is observed by \anubis in the ceiling configuration after the full selection procedure. This event would not be observed in the shaft configuration.
The display follows the same format as Figure~\ref{fig:ShaftConfigTS1}, except that the intersection points of the final-state particles in both RPC triplets in the ceiling Tracking Station are shown separately as individual tracking layers. These are displayed in the local curved coordinate system of the tracking layer (bottom).  
%Dashed lines represent the LLP trajectory, while solid lines show the trajectories of the ensuing charged final-state particles produced in the LLP decay. 
%Cross-sections of the ATLAS cavern and service shafts in the $zy$- (top left) and $xy$- (top right) planes are also shown, with dotted lines representing ATLAS' vertexing limit. 
%The intersection points of the final-state, charged particles with ANUBIS' tracking layers are shown for the ceiling using local curved coordinates (bottom), where each layer is a triplet of RPCs and the two layers are separated by $\sim1$ m.
%The intersection points are represented as crosses.
} 
\end{figure*}

Two configurations for the \anubis geometry and hence the observation of LLP decays are considered:
\begin{itemize}
%\item
%The `shaft-only' configuration follows the definition in the original ANUBIS proposal~\cite{bib:anubis_orig}, and considers only LLPs decaying inside the PX14 service shaft, beyond the first TS at the bottom of the shaft;
\item
The shaft configuration from Section~\ref{sec:ANUBIS_overview} corresponds to the `shaft+cone' scenario from the original \anubis proposal. 
It allows the LLP to decay within the PX14 shaft or within the UX15 \atlas experimental cavern provided the decay products traverse the lowest TS;
% effectively representing the full potential coverage of the shaft-only configuration.
\item
The ceiling scenario follows the ceiling configuration introduced in Section~\ref{sec:ANUBIS_overview} and described in detail in Section~\ref{sec:bkg_predictions}, albeit with conservative modifications for the signal case: 
the active decay volume starts 9.5~m away from the beamline.
This is motivated by the rapidly dropping density of material  in the barrel region of \atlas muon spectrometer at 4.5~m away from the beamline, to which 5~m are conservatively added. 
These additional 5~m account for the loss in the active decay volume from any material-dense regions in the \atlas muon spectrometer like support structures or active \atlas detector elements like RPCs or muon drift tubes, and a 30~cm safety margin around these regions on account of finite vertex resolution.
It should be noted that the \atlas vertexing limit is 7.5 m~\cite{ATLAS:2018tup}.
The active decay volume ends 1.5~m below the ceiling of the \atlas cavern, which accounts for the TS (1~m), fixings and services (0.2~m), and a safety margin (0.3~m) in front of the TS to ensure a reliable rejection of DVs that originate from the TS material. 
Hence, the radial length of the active decay volume of \anubis is shorter for the signal sensitivity projections than for background projections. 
%the LLPs must decay within the bounds of the ATLAS cavern between the vertexing limit of the ATLAS detector (radius = 7.5 m from the beamline) and the cavern walls.
\end{itemize}
Representative examples of in-cavern LLP decays for the shaft and ceiling configurations are shown in Figure~\ref{fig:ShaftConfigTS1} and Figure~\ref{fig:CeilingConfig}, respectively.

In the following, it is assumed that the \atlas magnetic field has negligible influence on the trajectory of final-state, charged particles for these configurations.
This is justified since the magnetic field will have a roughly isotropic effect on the jets of energetic charged particles from LLP decays, resulting in minimal impact on overall sensitivity.
In other words, the capability of measuring the transverse momenta of the charged particle tracks from LLP decays inside or close to the \atlas muon spectrometer and hence the invariant mass of the DV is conservatively neglected for the sensitivity projections in Section~\ref{sec:Sensitivity}. 
However, note that this experimental capability will allow \anubis+\atlas to provide a lower bound on the LLP mass from its decay into charged particles in case of a discovery.

\subsection{Sensitivity calculation} 
\label{sec:Sensitivity} 
This study probes values of $c\tau$ ranging from $10^{-3}$ to $10^{8}$~m that bound the expected sensitivity of \anubis, in increments of one-fifth of an order of magnitude.
%These bounds have been chosen as they encompass the projected sensitivity limits of the proposed ANUBIS detector configuration.
The geometrical and kinematic acceptance of \anubis, $\acc$, is calculated by applying the selections from Section~\ref{sec:SenEventSelec} to MC simulations from Section~\ref{sec:SenSimulations}, counting the number of DVs that are reconstructed in a given detector configuration outlined in Section~\ref{sec:SenGeometricModel}, and dividing by the total number of DVs produced.
The hit detection efficiency is assumed to be 100\% in the \acc calculation, which is a good approximation as motivated in Section~\ref{sec:bkg_signatures}.
The selection efficiency for DVs from signal events $\eps_{\rm sig}$ accounts for the efficiency of the discrimination between DVs from signal and background events considering vertex-level observables like the hit multiplicity.
The \atlas LLP search features $62\%<\eps_{\rm sig}^{\rm ATLAS}<74\%$ for $15<m_s/\GeV<60$ ~\cite{ATLAS:2018tup}. 
Hence, a value of $\eps_{\rm sig}^{\rm ANUBIS}=50\%$ is conservatively assumed for the \anubis sensitivity projections, motivated by the similarity in the detector technology between \anubis and the barrel region of the \atlas muon spectrometer.
Note that for the background estimate in Section~\ref{sec:bkg_predictions} a two times higher selection efficiency was conservatively assumed for \anubis compared to \atlas, 
despite the conservative assumption $\eps_{\rm sig}^{\rm ANUBIS} < \eps_{\rm sig}^{\rm ATLAS}$ above.

%The potential detection efficiency and acceptance of ANUBIS are calculated by determining the ratio of observed decays $N_\text{obs}$ to the total number of simulated LLP production events $N_\text{tot}$ for a given $\tau$.
%This ratio, denoted as $\frac{N_\text{obs}}{N_\text{tot}}$, represents ANUBIS’ acceptance multiplied by its expected efficiency for detecting LLPs in a given configuration.
%The detection efficiency is assumed to be 100\% as outlined in Section~\ref{sec:bkg_signatures}. 
%The selection efficiency for DVs from signal events is assumed as $\eps_{\rm sig}=30\%$ motivated by Refs~\cite{ATLAS:2018tup,ATLAS:2025pak}, which accounts for the need to discriminate between DVs from signal and background using machine learning tools on observables like the hit multiplicity.
%%The detector hardware and reconstruction efficiencies are assumed to be 100\%, though exact values could be included in future through additional efficiency terms after further studies.

The number of signal events $N_\text{sig}$ expected to be observable is given by
\begin{equation} N_\text{sig} = \mathcal{L}_\text{HL-LHC} \cdot \sigma_{h} \cdot 2 \cdot \br(h \rightarrow ss) \cdot \acc \cdot \eps_{\rm sig}\,,
\label{eq:sensitivity}
\end{equation} 
where $\mathcal{L}_\text{HL-LHC}=3~\invab$ is the integrated luminosity of $pp$ collisions at $\sqrt{s}=14~\TeV$ at the HL-LHC, $\sigma_{h}=54.67\text{ pb}$~\cite{bib:pdg2024} represents the Higgs boson ggF production cross section at $\sqrt{s}=14$ TeV, and $\br(h \to ss)$ denotes the exotic branching ratio of the  125~GeV Higgs boson into an $ss$ pair.
The factor of 2 features in Eq.~\eqref{eq:sensitivity} since a pair of LLPs are produced in the Higgs decay, either of which can potentially provide a reconstructable DV.

In practice, the expected sensitivity of \anubis to an exotic 125 GeV Higgs boson decay into an $ss$ pair is calculated by solving Eq.~\eqref{eq:sensitivity} for $\br(h \to ss)$ after determining \acc for a given $m_s$ and $c\tau$ hypothesis
%using MC simulations following Section~\ref{sec:SenGeometricModel} 
and setting $N_\text{sig}$ to the number of events required for signal observation in a given detector configuration. 

Previously reported sensitivity for \anubis under a conservative background assumption~\cite{bib:anubis_orig,PBC:2025sny,Brandt:2025fdj,satterthwaite_2022_292b9-eck73} was given using the $5\sigma$ discovery as a figure of merit. 
Following the formalism from Ref.~\cite{CowanStat}, this corresponds to values of $N_{\text{sig}}\geq102$, $N_{\text{sig}}\geq45$, and $N_{\text{sig}}\geq52$ using Eqs.~\eqref{eq:bgnum}--\eqref{eq:bgnumshaft} for the ceiling conservative, nominal, and shaft configurations, respectively. 
However, for better comparison with other LLP experiments and existing experimental limits, signal thresholds were also determined for an exclusion sensitivity at the 95\% confidence level (CL) using the CLs method~\cite{Read:2002hq}. 
These thresholds for the \anubis configurations are $N_{\text{sig}}\geq36$, $N_{\text{sig}}\geq17$, and $N_{\text{sig}}\geq19$ assuming background estimates from Eqs.~\eqref{eq:bgnum}--\eqref{eq:bgnumshaft}, respectively. 
The conservative background sensitivity projections will use the 95\% CL exclusion sensitivity unless otherwise stated. 
The background-free sensitivity projections require $N_\text{sig}\geq4$ for all detector configurations.

\subsection{Results} 
\label{sec:SensitivityResults}

Throughout this section the 1DV selection limits will be considered unless otherwise stated, as this is the most conservative scenario. Likewise, the 95\% CL exclusion sensitivity will be provided.

\begin{figure*}[tb]
\centering
\begin{subfigure}[b]{0.45\textwidth}
\includegraphics[width=0.9\textwidth]{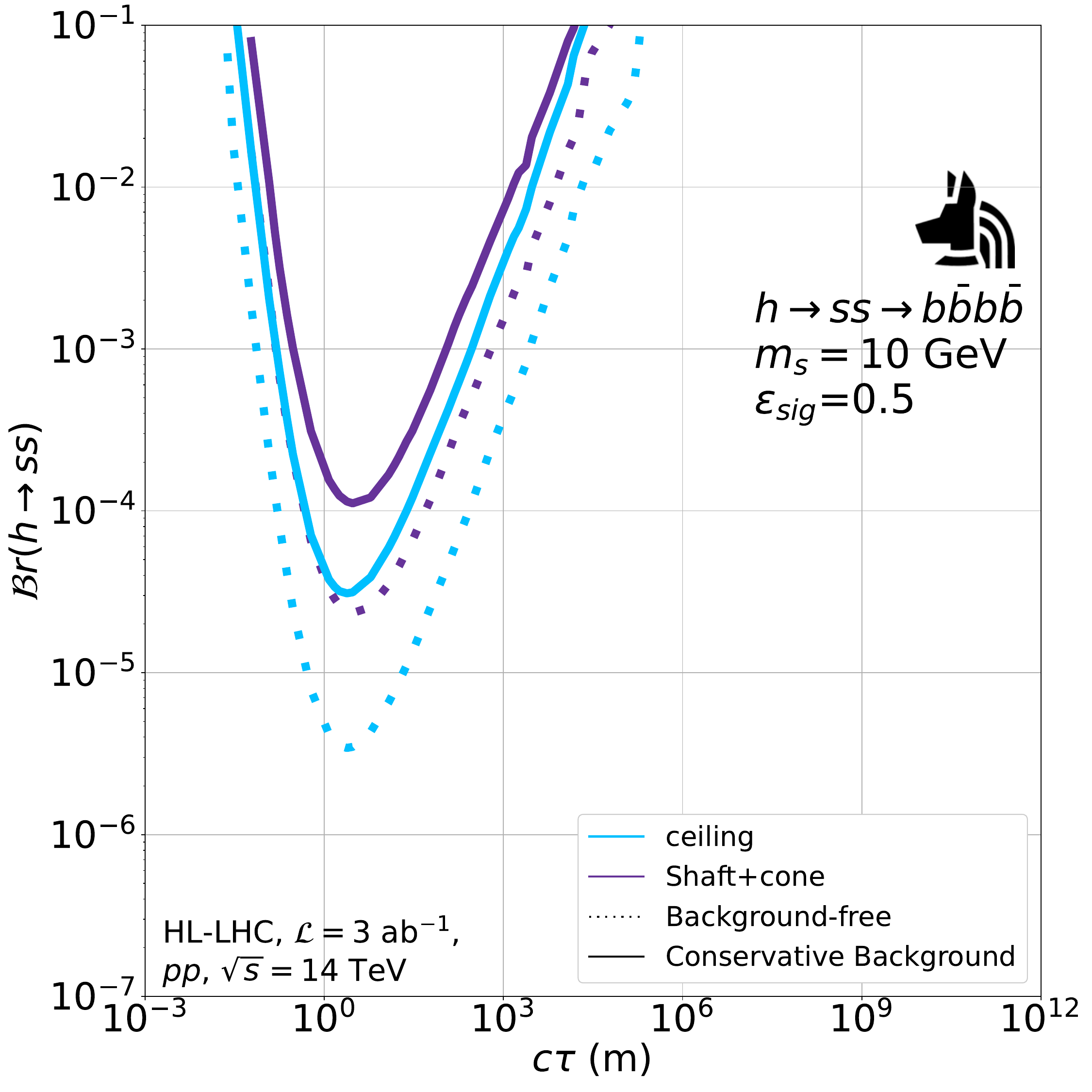}
\caption{}
\end{subfigure}
\begin{subfigure}[b]{0.45\textwidth}
\includegraphics[width=0.9\textwidth]{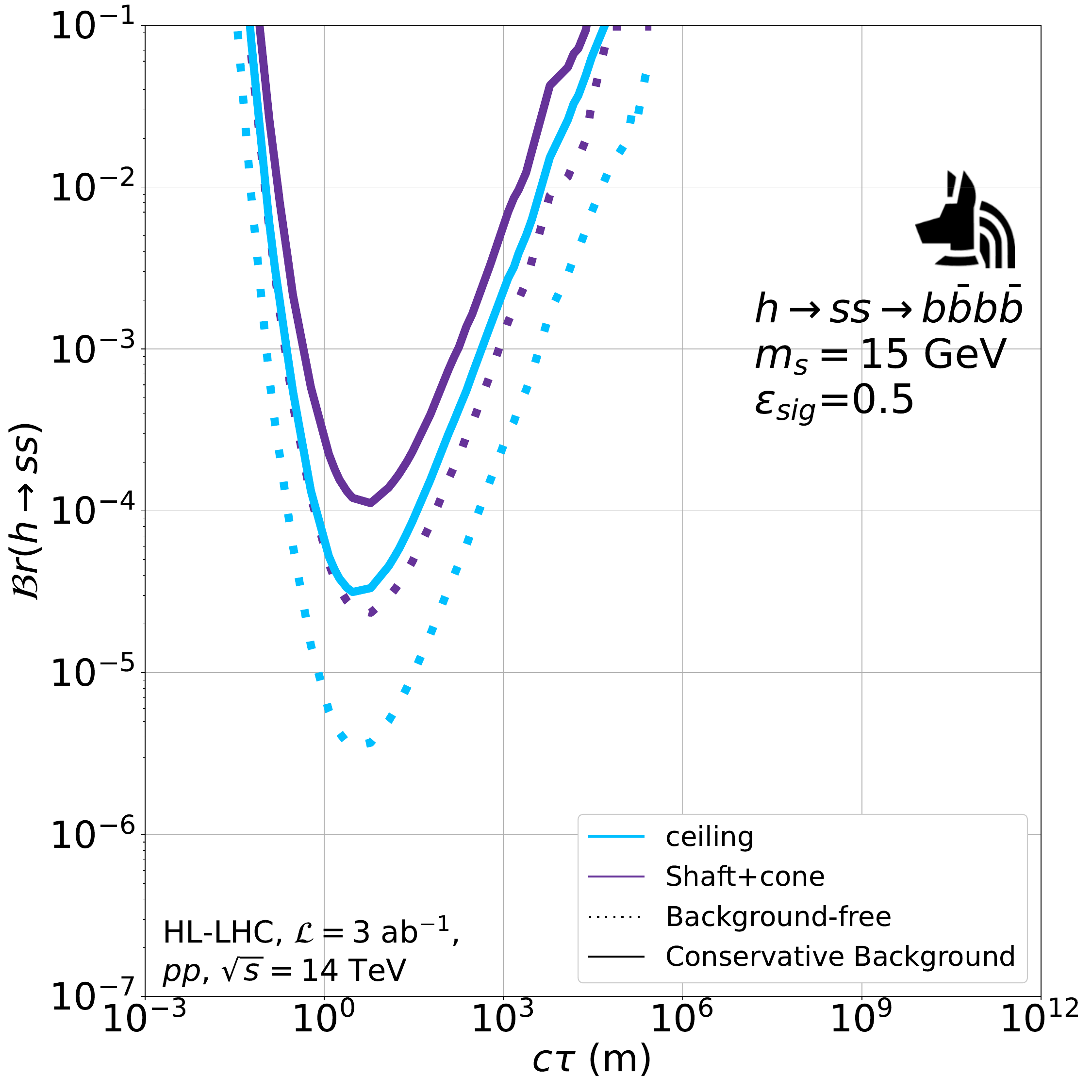}
\caption{}
\end{subfigure}
\\
\begin{subfigure}[b]{0.45\textwidth}
\includegraphics[width=0.9\textwidth]{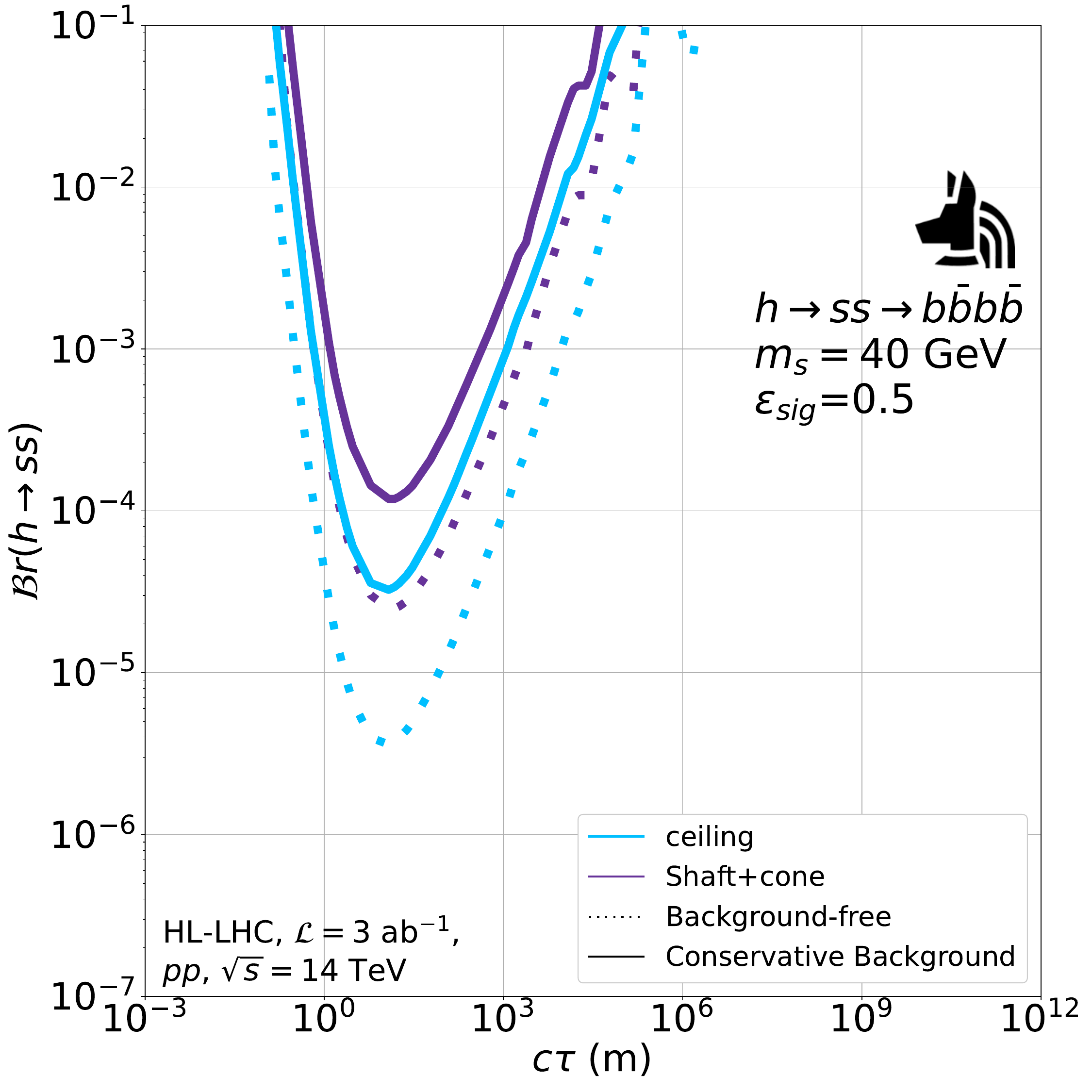}
\caption{}
\end{subfigure}
\begin{subfigure}[b]{0.45\textwidth}
\includegraphics[width=0.9\textwidth]{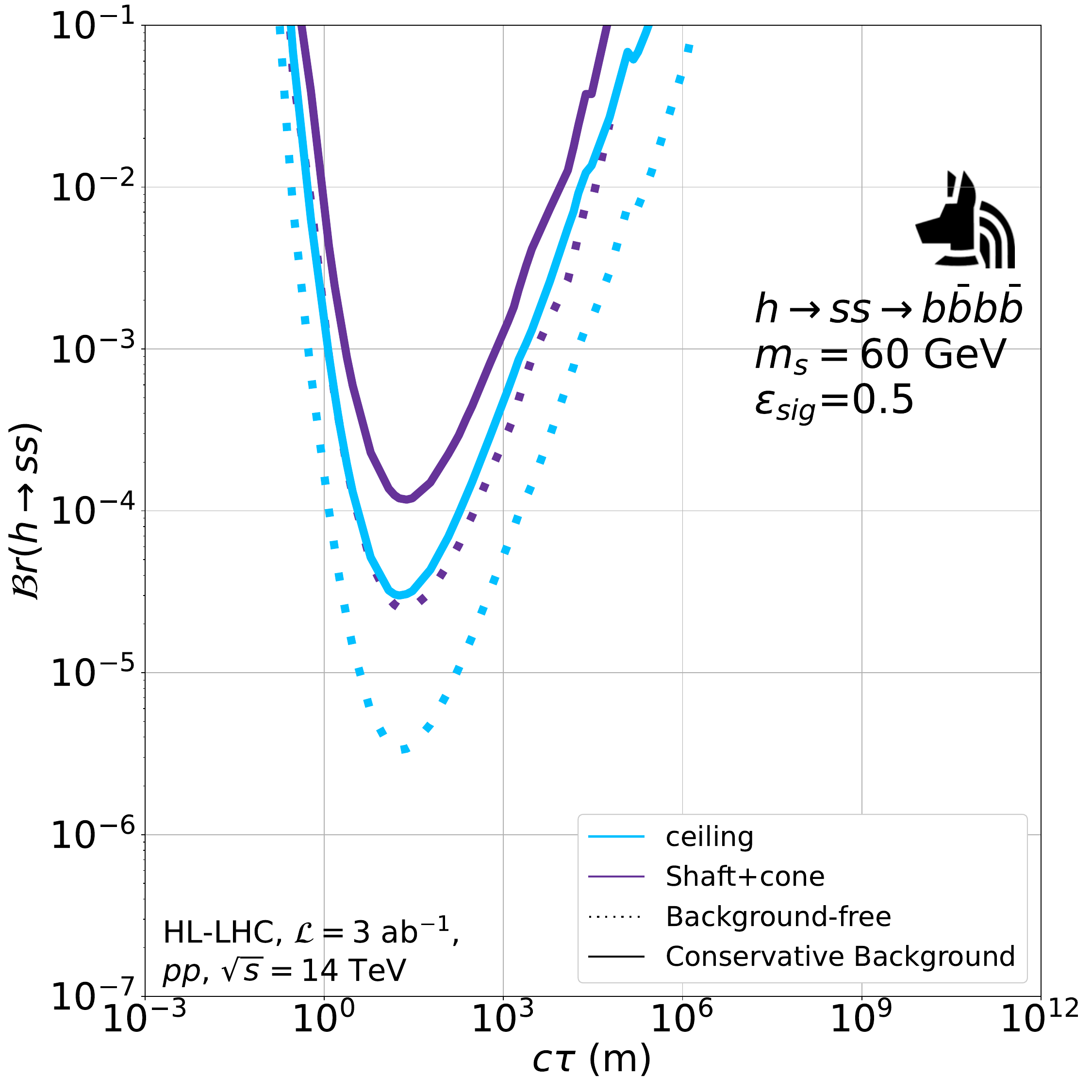}
\caption{}
\end{subfigure}
\caption{
\label{fig:SensitivityAll}
%\vspace{-5mm}
The 95\% CL exclusion sensitivity of \anubis to an exotic branching ratio of the 125~GeV Higgs boson into a pair of long-lived scalars $s$ with masses of (a)~10~GeV, (b)~15~GeV, (c)~40~GeV, and (d)~60~GeV using 3~\invab of $pp$ collisions at the HL-LHC, for the ceiling~(blue) and shaft~(purple) configurations of the \anubis detector, with the 1DV selection.
The projections are shown for the conservative (solid lines) and background-free (dotted lines) scenarios, which bracket the expected sensitivity of \anubis.
}
\end{figure*}

The expected sensitivity of \anubis to an exotic branching ratio of the 125~GeV Higgs boson into a pair of scalar LLPs, \ie $\br(h\to ss)$,  using 3~\invab of $pp$ collisions at the HL-LHC is presented in Figure~\ref{fig:SensitivityAll} for $m_s=10,15,40,$ and 60~GeV for the 1DV selection.
The results are shown for the two configurations of the \anubis detector: shaft and ceiling.
Both conservative projections and background-free projections that bracket the expected sensitivity of \anubis are provided.
% The fluctuations at $c\tau\gtrsim10^6$ are due to the finite number of simulated MC events in this region.
% They have no physics impact given that this region is expected to be probed by searches for Higgs boson decays into invisible particles $h\to\text{invisible}$~\cite{Cepeda:2019klc}, since the majority of LLP decays will happen outside of the ATLAS (or CMS) detector volume and hence be registered as \met in this large $c\tau$ regime.

The ceiling configuration provides the best sensitivity and can probe $\br(h\to ss)$ down to $\mathcal{O}(10^{-5})$ and $\mathcal{O}(10^{-6})$ for the conservative and  background-free projections, respectively.
It is followed by the shaft+cone configuration, where the sensitivity to $\br(h\to ss)$ is reduced by about an order of magnitude in both the background-free and conservative projections. If only decays in the PX14 shaft are considered the sensitivity is further reduced by a similar amount.
This sensitivity ranking is expected from the $\leff$, 
%\ie the integral of the solid angle and the path element  over the active detector volume, 
which is largest for the ceiling configuration.
Combined with fewer practical constraints outlined in Section~\ref{sec:ANUBIS_overview}, this confirms the choice of the ceiling configuration as the default layout of the future \anubis detector.

% A detailed inspection of Figure~\ref{fig:SensitivityAll} reveals several subtle aspects. 
% The sensitivity of the shaft+cone and the shaft-only configuration are closer to each other for large $c\tau \gtrsim 10^3~\metre$ than at about $c\tau\approx10~\metre$ that achieves the highest sensitivity.
% %, where the shaft-only configuration lags behind by up to one order of magnitude.
% This can be understood since the sensitivity scales with \leff in the large $c\tau$ limit, whereas at smaller $c\tau$ the exponential nature of LLP decays comes into play, giving preference to the shaft+cone configuration over the shaft-only configuration, since the former features an active decay volume that is closer to the interaction point.
A subtle aspect in Figure~\ref{fig:SensitivityAll} is the shift of the $c\tau$ value where the best sensitivity is achieved from $c\tau=2.4$~m for $m_s=10~$GeV to 18~m for $m_s=60~\GeV$ considering the default ceiling configuration, which is reflected in similar trends for the other configurations. 
This effect is due to the increase of the boost $\gamma$ with decreasing $m_s$.

\enlargethispage{1cm}
\begin{figure*}[tb]
\centering
\begin{subfigure}[b]{0.4\textwidth}
\includegraphics[width=\textwidth]{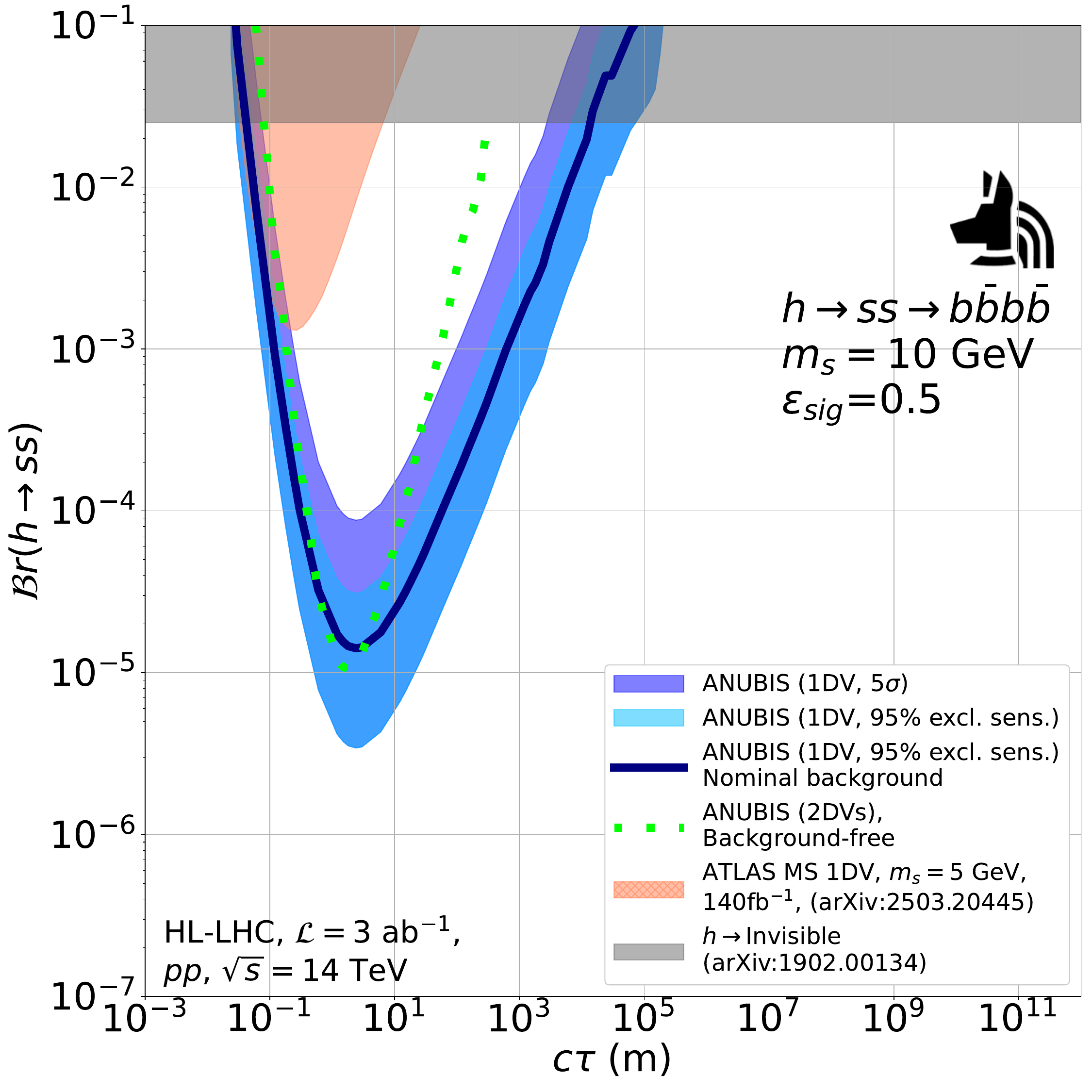}
\caption{}
\end{subfigure}
\qquad
\begin{subfigure}[b]{0.4\textwidth}
\includegraphics[width=\textwidth]{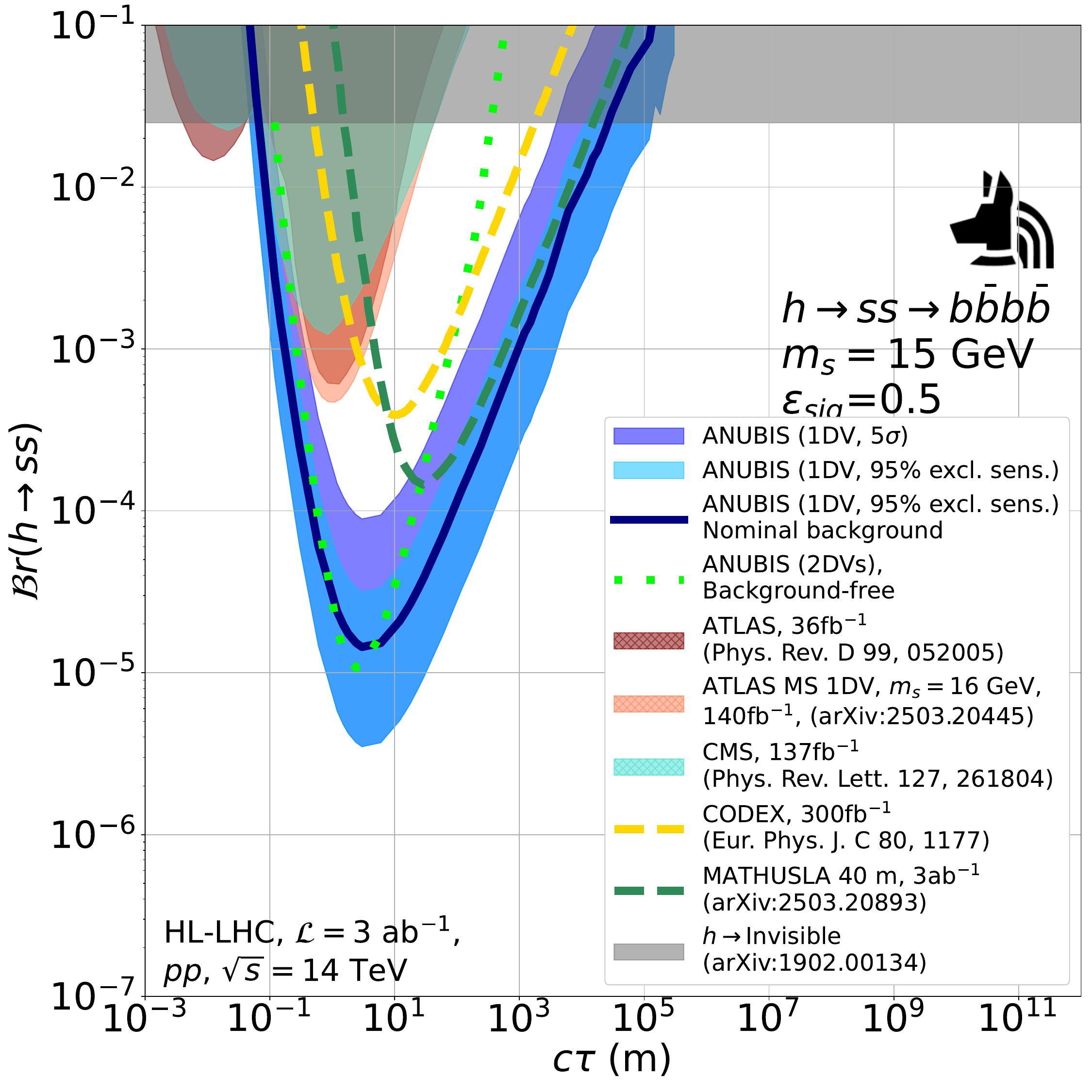}
\caption{}
\end{subfigure}
\\
\begin{subfigure}[b]{0.4\textwidth}
\includegraphics[width=\textwidth]{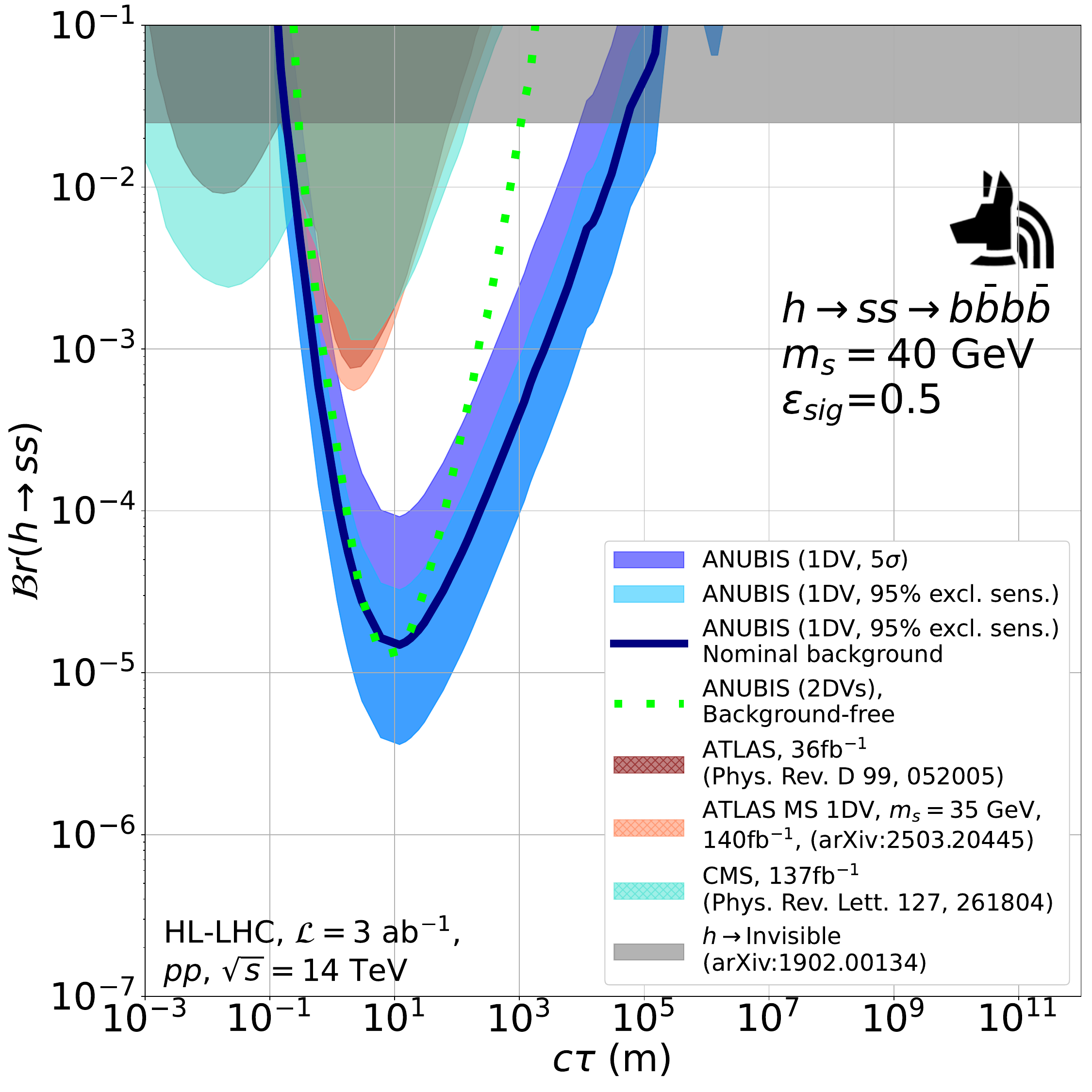}
\caption{}
\end{subfigure}
\qquad
\begin{subfigure}[b]{0.4\textwidth}
\includegraphics[width=\textwidth]{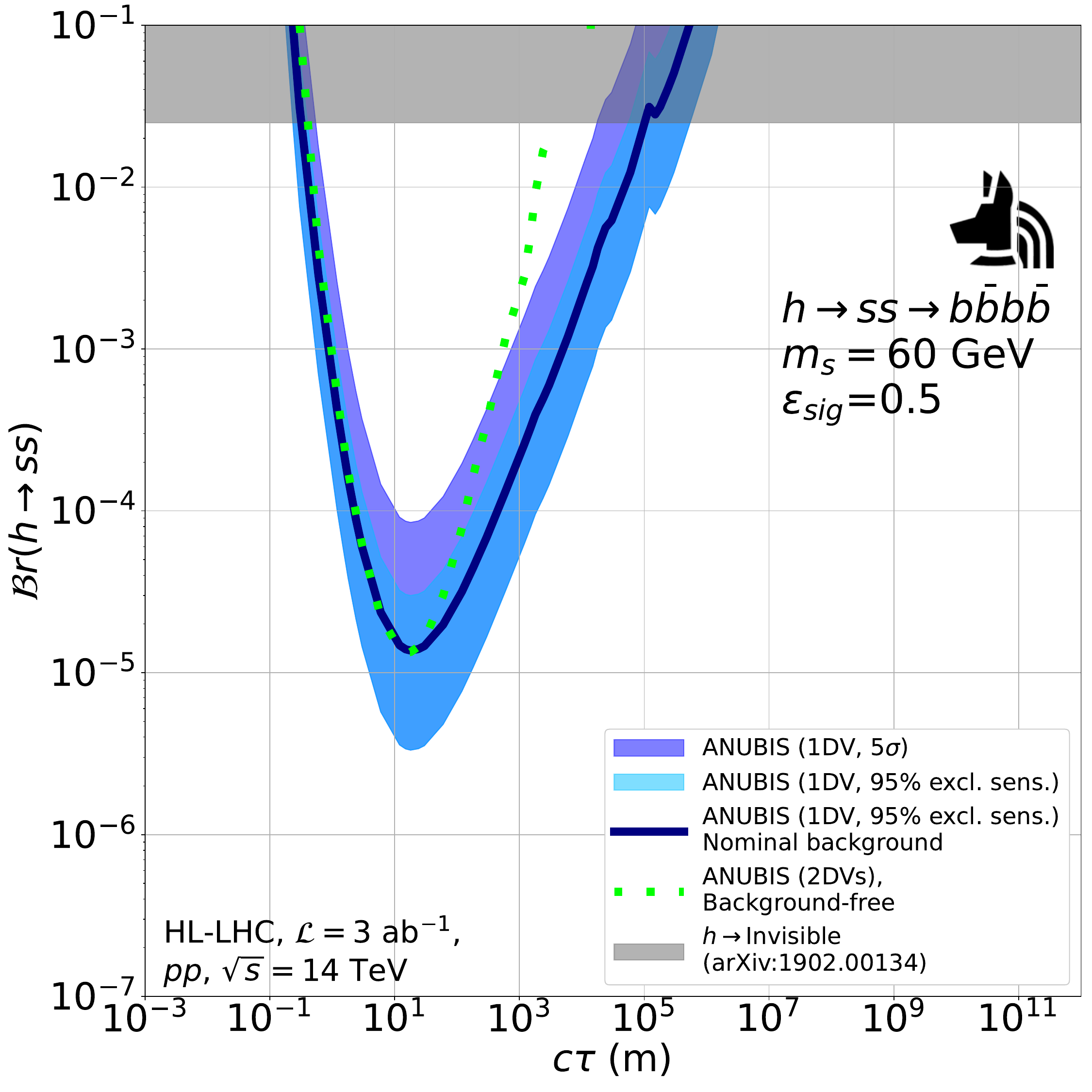}
\caption{}
\end{subfigure}
%\end{center}
\caption{
\label{fig:SensitivityCeilingOnly}
The sensitivity of \anubis to an exotic branching ratio of the 125~GeV Higgs boson into a pair of long-lived scalars $s$ with masses of (a)~10~GeV, (b)~15~GeV, (c)~40~GeV, and (d)~60~GeV using 3~\invab of $pp$ collisions at the HL-LHC. 
%The ANUBIS projections are shown for the default ceiling configuration considering the conservative~(solid lines) and background-free~(dotted lines) scenarios, which bracket the expected sensitivity.
The \anubis projections are shown as a band, where the lower (upper) edge corresponds to the background-free (conservative background) scenario for the 1DV selection. The nominal background scenario for the 1DV selection is shown as a solid line. The 5$\sigma$ discovery and 95\% CL exclusion limits are both shown, where they have a common lower edge. The background-free \anubis projection for the 2DVs selection is shown as a dotted line.
For comparison, the sensitivity projections from other proposed transverse detectors \codex~\cite{CODEX-b:2019jve},
%MATHUSLA 200~m~\cite{ Chou:2016lxi}
and \mathusla 40~m~\cite{MATHUSLA:2025zyt} at the HL-LHC are also presented, alongside the projected sensitivity from $h\to\text{invisible}$ decays at the HL-LHC~\cite{Cepeda:2019klc}. 
The latest LHC results from \atlas~\cite{ATLAS:2018tup,ATLAS:2025pak} and \cms~\cite{CMS:2021juv} are also shown for reference.
}
\end{figure*}

Figure~\ref{fig:SensitivityCeilingOnly} compares the projected \anubis discovery and exclusion sensitivities in the default ceiling configuration to background-free exclusion projections from other major proposed transverse experiments at the HL-LHC: \codex~\cite{CODEX-b:2019jve} and \mathusla~\cite{MATHUSLA:2025zyt}.
%MATHUSLA~\cite{Chou:2016lxi,MATHUSLA:2025zyt}.
Also shown are the exclusion limits at 95\% CL from existing \atlas~\cite{ATLAS:2018tup,ATLAS:2025pak} and \cms~\cite{CMS:2021juv} searches, since no HL-LHC projections from these collaborations exist for this signature.
The projected HL-LHC exclusion limits at 95\% CL from searches for Higgs boson decays into invisible particles $h\to\text{invisible}$~\cite{Cepeda:2019klc} are also shown for reference.
The \anubis sensitivity is represented as a band where the lower (upper) edge corresponds to the background-free (conservative background) scenario. This highlights the range of potential branching ratios that could be probed between the best-case scenario and the expected conservative background level. 
%, since in the large $c\tau$ regime considered in this document the majority of the LLP decays will happen outside of the ATLAS or CMS detector volume and hence be registered as \met.
%In this figure, there is a notable variation observed in the high-$c\tau$ region, which arises from statistical fluctuations due to the low number of events intersecting ANUBIS at these lifetimes.
%However, they lie within the exclusion limits for the Higgs boson branching ratio to invisible particles, rendering ANUBIS' sensitivity at these values insignificant.

At its peak sensitivity at $c\tau\approx12~\metre$, \anubis demonstrates an improvement in exclusion sensitivity of an exotic $\br(h\to ss)$ by two to three orders of magnitude compared to the latest \atlas and \cms results.
%, which searched for a 40 GeV BSM LLP by identifying displaced vertices within the muon spectrometer~\cite{ATLAS:2018tup}.
Moreover, for $\br(h\to ss)=10^{-3}$, \anubis extends the $c\tau$ reach of \atlas and \cms by three~(four) orders of magnitude considering the conservative~(background free) projections, and can uniquely probe $c\tau$ values up to $10^3~\metre$~($10^{4}~\metre$).
Considering $\br(h\to ss)=10^{-2}$, \anubis' $c\tau$ reach of $10^4~\metre$~($10^{5}~\metre$) for conservative~(background-free) projections approaches within three~(two) orders of magnitude the limit set by Big Bang nucleosynthesis constraints~\cite{Fradette:2017sdd}.
Compared to \atlas and \cms sensitivity projections at the HL-LHC discussed within the PBC group~\cite{bib:curtin_hllhc}, \anubis provides one~(two) orders of magnitude improvement in sensitivity to an exotic $\br(h\to ss)$ at $c\tau=10^{3}$~m, and extends the $c\tau$ reach by more than one~(two) orders of magnitude for conservative~(background-free) projections.
%compared to the ATLAS HL-LHC projection.
%The CMS HL-LHC projection covers a similar $c\tau$ range but has approximately an order of magnitude better sensitivity than the ATLAS projection.

Comparing the transverse experiments for the HL-LHC phase, both \text{\codex} and \anubis reach their maximum sensitivity at $c\tau\approx10~\metre$. 
Considering conservative~(background-free) projections, \anubis surpasses \codex by about one~(two) orders of magnitude in terms of sensitivity to $\br(h\to ss)$ across the entire $c\tau$ range for $m_s=15$~GeV. 
Similarly, \anubis' conservative limits surpass the \mathusla~40~m projections for $c\tau\lesssim50~\metre$ for $m_s=15$~GeV by up to an order of magnitude, and \anubis' background-free projections surpass them across the whole $c\tau$ range and by up to three orders of magnitude for $c\tau\approx 5~\metre$. 
%The MATHUSLA~40~m projections from Ref.~\cite{MATHUSLA:2025zyt} use the exclusion from a background-free search and a 49\% signal reconstruction efficiency.%

In terms of the $c\tau$ reach and taking $\br(h\to ss)=10^{-3}$ as a benchmark, \anubis surpasses \codex by two~(three) orders of magnitude and \mathusla~40~m by one~(two) orders of magnitude for conservative~(background-free) projections for $m_s=15$ GeV.
Note that \anubis has the unique advantage among the transverse experiments, featuring a continuous active decay volume that, in combination with \atlas, extends all the way to the interaction point. 
This allows for seamless coverage across the entire $c\tau$ range.

\begin{figure*}[tb]
\centering
\begin{subfigure}[b]{0.4\textwidth}
\includegraphics[width=\textwidth]{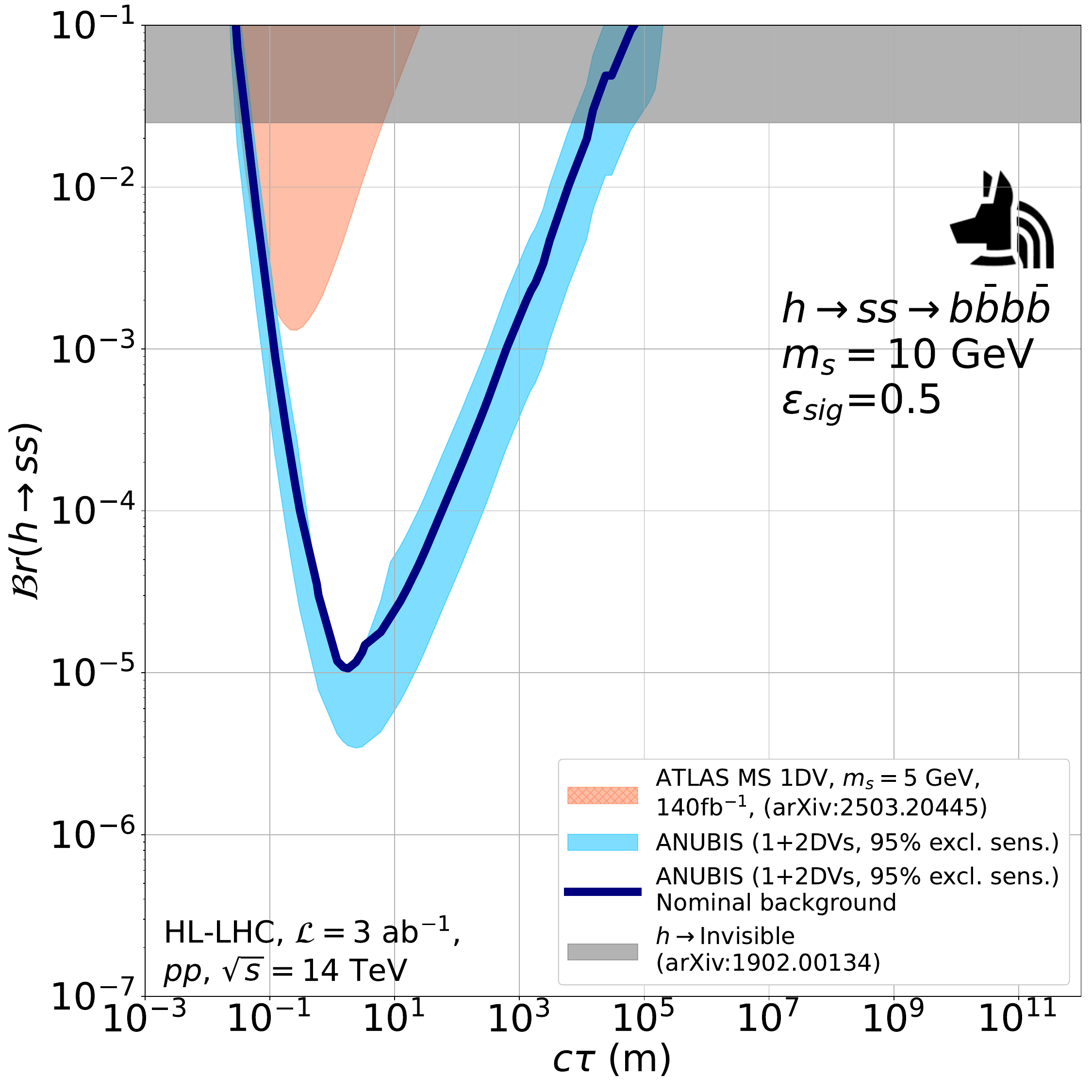}
\caption{}
\end{subfigure}
\qquad
\begin{subfigure}[b]{0.4\textwidth}
\includegraphics[width=\textwidth]{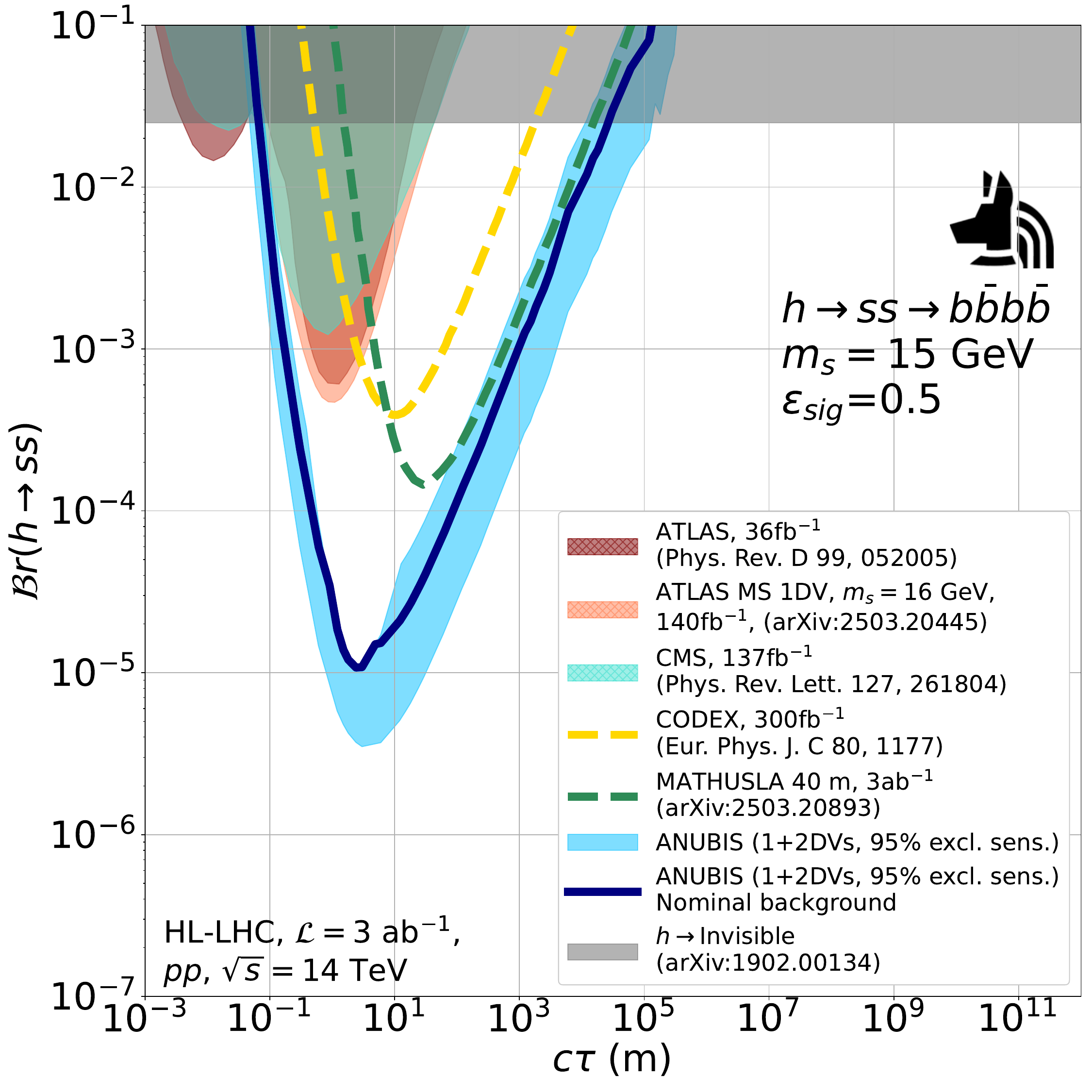}
\caption{}
\end{subfigure}
\\
\begin{subfigure}[b]{0.4\textwidth}
\includegraphics[width=\textwidth]{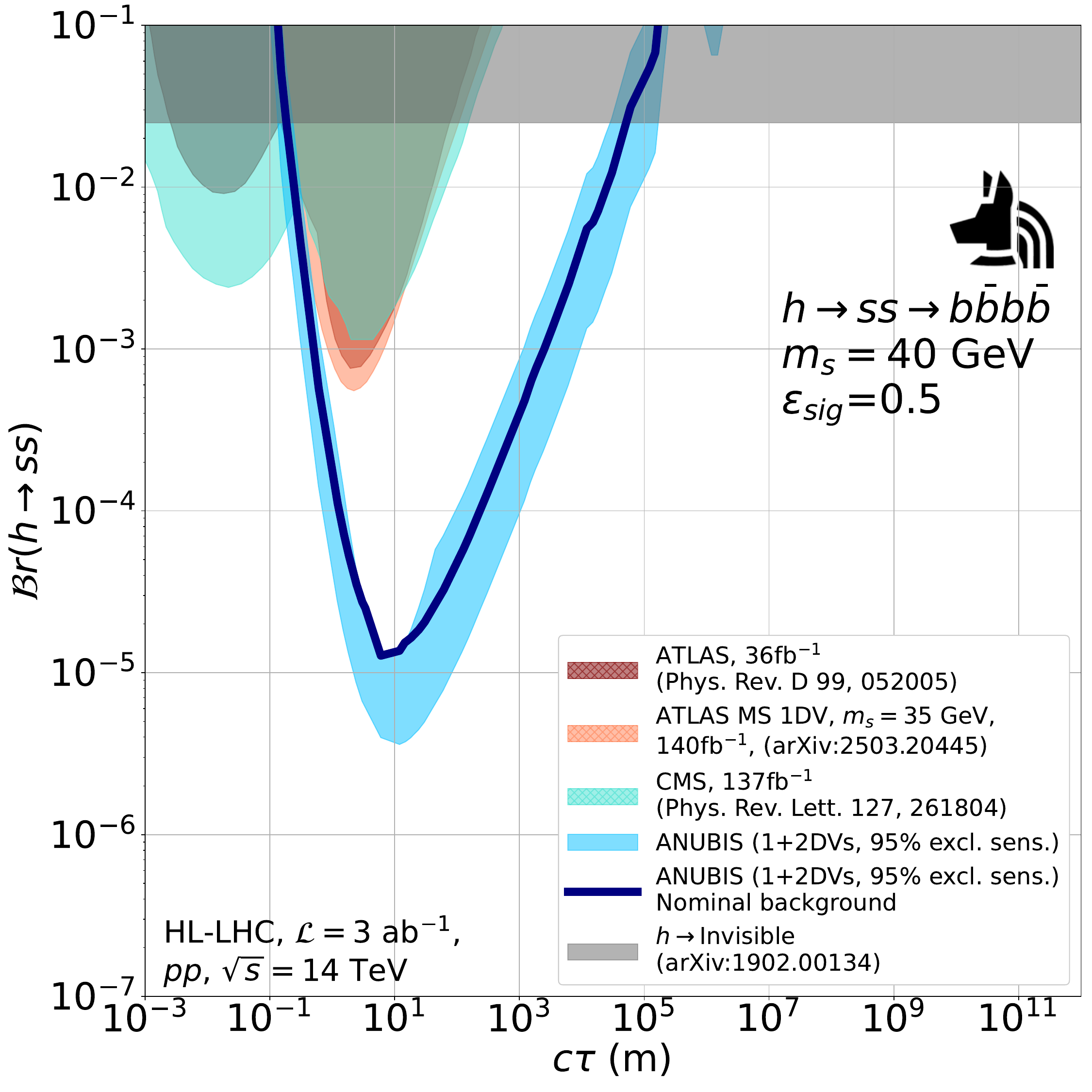}
\caption{}
\end{subfigure}
\qquad
\begin{subfigure}[b]{0.4\textwidth}
\includegraphics[width=\textwidth]{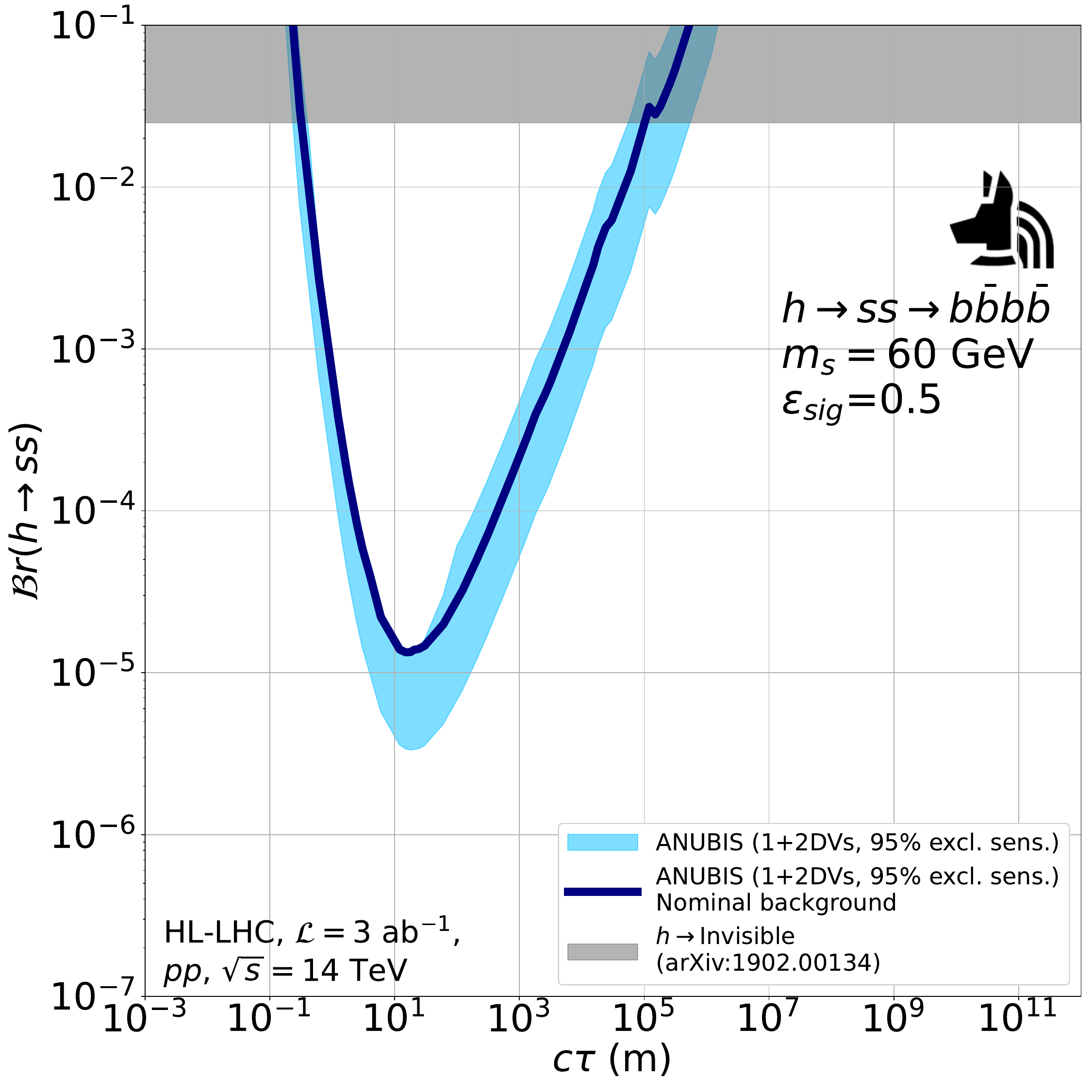}
\caption{}
\end{subfigure}
\caption{
\label{fig:SensitivityCeilingOnlyCombined}
The 95\% CL exclusion sensitivity of \anubis to an exotic branching ratio of the 125~GeV Higgs boson into a pair of long-lived scalars $s$ with masses of (a)~10~GeV, (b)~15~GeV, (c)~40~GeV, and (d)~60~GeV using 3~\invab of $pp$ collisions at the HL-LHC. 
%The ANUBIS projections are shown for the default ceiling configuration considering the conservative~(solid lines) and background-free~(dotted lines) scenarios, which bracket the expected sensitivity.
The combined single and two displaced vertices projections for \anubis are shown as a band, the lower (upper) edge being the minimum of the background-free (conservative) 1DV and the 2DVs limits. The nominal background scenario for the 1DV selection is shown as a solid line.
For comparison, the sensitivity projections from other proposed transverse detectors \codex~\cite{CODEX-b:2019jve}, 
%MATHUSLA 200~m~\cite{ Chou:2016lxi} 
and \mathusla 40~m~\cite{MATHUSLA:2025zyt} at the HL-LHC are also presented, alongside the projected sensitivity from $h\to\text{invisible}$ decays at the HL-LHC~\cite{Cepeda:2019klc}. 
The latest LHC results from \atlas~\cite{ATLAS:2018tup,ATLAS:2025pak} and \cms~\cite{CMS:2021juv} are also shown for reference.
}
\end{figure*}

These sensitivity limits can be further improved using the 2DVs selection, which is effectively background-free. The combination of the 1DV and 2DVs limits then represents the optimised reach that \anubis would have to this model. Figure~\ref{fig:SensitivityCeilingOnlyCombined} shows this band, with the lower (upper) edge being the minimum of the background-free (conservative) 1DV and the 2DVs limits. This improves the maximum sensitivity in the conservative background scenario by up to half an order of magnitude, and gives a narrower band below $c\tau=\mathcal{O}(10)~\metre$.\\ 

The comparisons in terms of the sensitivity to $\br(h\to ss)$ are summarised in Table~\ref{tab:Br_Sensitivity} for four representative benchmarks with $m_s=15~\GeV$ ($c\tau = 3,~100~\metre$) and $m_s=40~\GeV$ ($c\tau = 10,~100~\metre$).
The $c\tau$ reach for $\br(h\to ss)=10^{-3}$ is shown in Table~\ref{tab:ctau_reach_II} for  $m_s=15$ and 40~GeV.
In combination with Figure~\ref{fig:SensitivityCeilingOnly}, both tables indicate that transverse experiments are an indispensable component of the experimental landscape of LLP searches at the HL-LHC, significantly extending the reach of existing experiments.
Furthermore, the results underline that \anubis is a competitive contender among the transverse detector initiatives.
%In the most stringent configuration of in-shaft decay searches, ANUBIS demonstrates a sensitivity improvement of approximately three-quarters of an order of magnitude over CODEX-b when targeting four observations of a 10 GeV Higgs-portal LLP with a $c\tau$ of 3 m.
%Furthermore, the project's shaft configuration yields an enhancement exceeding one order of magnitude in sensitivity relative to the CODEX-b projections for the same LLP model, with the ceiling configuration further increasing sensitivity by an additional three-quarters of an order of magnitude.

\begin{table*}[tb]
\centering
\small
\caption{
The exclusion sensitivity of \anubis in the default ceiling configuration to $\br(h\to ss)$ for $m_s=15~\GeV$ with $c\tau=3$ and 100~m, and $m_s=40~\GeV$ with $c\tau=10$ and 100~m, using 3~\invab of $pp$ collisions at the HL-LHC.
Both conservative~(conserv.) and background-free (bg.-free) projections are shown, which bracket the expected \anubis sensitivity.
For comparison, the sensitivity projections from other proposed transverse detectors \codex~\cite{CODEX-b:2019jve}, \mathusla 200~m~\cite{Chou:2016lxi} and \mathusla 40~m~\cite{MATHUSLA:2025zyt} at the HL-LHC are also presented, alongside the latest LHC results from \atlas~\cite{ATLAS:2025pak} and \cms~\cite{CMS:2021juv} using $\sim140$~\invfb of $pp$ collisions.
}

\renewcommand{\arraystretch}{1.1}
\begin{tabular}{l|cccc}
\hline
\hline
\hbox{}\hspace{27mm}$\boldsymbol{m_s}$ & \textbf{15 GeV} & \textbf{15 GeV} & \textbf{40 GeV} & \textbf{40 GeV} \\
\hbox{}\hspace{27mm}\textbf{c$\boldsymbol\tau$} & \textbf{3 m} & \textbf{100 m} & \textbf{10 m} & \textbf{100 m} \\
\hline
\textbf{\atlas, 140~\invfb} & 1.1 $\times$ 10$^{-3}$ & 1.8 $\times$ 10$^{-1}$ & 1.8 $\times$ 10$^{-3}$ & 3.4 $\times$ 10$^{-2}$ \\
\textbf{\cms, 137~\invfb} & 2.4 $\times$ 10$^{-3}$ & 6.3 $\times$ 10$^{-2}$ & 1.8 $\times$ 10$^{-3}$ & 1.5 $\times$ 10$^{-2}$ \\
\textbf{\codex} & 8.0 $\times$ 10$^{-4}$ & 1.6 $\times$ 10$^{-3}$ & - & - \\
\textbf{\mathusla 40} & 3.7 $\times$ 10$^{-3}$ & 2.4 $\times$ 10$^{-4}$ & - & - \\
% \textbf{ANUBIS 1DV 5$\sigma$ conserv.} & 8.9 $\times$ 10$^{-5}$ & 7.1 $\times$ 10$^{-4}$ & 9.4 $\times$ 10$^{-5}$ & 3 $\times$ 10$^{-4}$ \\
% \textbf{ANUBIS 1DV 5$\sigma$ bg.-free} & 3.5 $\times$ 10$^{-6}$ & 2.8 $\times$ 10$^{-5}$ & 3.7 $\times$ 10$^{-6}$ & 1.2 $\times$ 10$^{-5}$ \\
% \textbf{ANUBIS 2DVs $5\sigma$ conserv.} & 2.8 $\times$ 10$^{-4}$ & 3.8 $\times$ 10$^{-2}$ & 3.4 $\times$ 10$^{-4}$ & 5.7 $\times$ 10$^{-3}$ \\
% \textbf{ANUBIS 2DVs $5\sigma$ bg.-free} & 1.1 $\times$ 10$^{-5}$ & 1.5 $\times$ 10$^{-3}$ & 1.3 $\times$ 10$^{-5}$ & 2.2 $\times$ 10$^{-4}$ \\
% \textbf{ANUBIS 1+2DVs $5\sigma$ conserv.} & 1.1 $\times$ 10$^{-5}$ & 7.1 $\times$ 10$^{-4}$ & 1.3 $\times$ 10$^{-5}$ & 2.2 $\times$ 10$^{-4}$ \\
% \textbf{ANUBIS 1+2DVs $5\sigma$ bg.-free} & 3.5 $\times$ 10$^{-6}$ & 2.8 $\times$ 10$^{-5}$ & 3.7 $\times$ 10$^{-6}$ & 1.2 $\times$ 10$^{-5}$ \\
\textbf{\anubis 1DV CLs conserv.} & 3.2 $\times$ 10$^{-5}$ & 2.5 $\times$ 10$^{-4}$ & 3.3 $\times$ 10$^{-5}$ & 1.0 $\times$ 10$^{-4}$ \\
\textbf{\anubis 1DV CLs bg.-free} & 3.5 $\times$ 10$^{-6}$ & 2.8 $\times$ 10$^{-5}$ & 3.7 $\times$ 10$^{-6}$ & 1.2 $\times$ 10$^{-5}$ \\
\textbf{\anubis 2DVs CLs conserv.} & 9.8 $\times$ 10$^{-5}$ & 1.3 $\times$ 10$^{-2}$ & 1.2 $\times$ 10$^{-4}$ & 2.0 $\times$ 10$^{-3}$ \\
\textbf{\anubis 2DVs CLs bg.-free} & 1.1 $\times$ 10$^{-5}$ & 1.5 $\times$ 10$^{-3}$ & 1.3 $\times$ 10$^{-5}$ & 2.2 $\times$ 10$^{-4}$ \\
\textbf{\anubis 1+2DVs CLs conserv.} & 1.1 $\times$ 10$^{-5}$ & 2.5 $\times$ 10$^{-4}$ & 1.3 $\times$ 10$^{-5}$ & 1.0 $\times$ 10$^{-4}$ \\
\textbf{\anubis 1+2DVs CLs bg.-free} & 3.5 $\times$ 10$^{-6}$ & 2.8 $\times$ 10$^{-5}$ & 3.7 $\times$ 10$^{-6}$ & 1.2 $\times$ 10$^{-5}$ \\
\hline
\hline
\end{tabular}
\label{tab:Br_Sensitivity}
\end{table*}

\begin{table*}[tb]
\centering
\small

\caption{
The expected $c\tau$ reach of \anubis in the default ceiling configuration to an $\br(h\to ss)=0.001$ for $m_s = 15$ and $40~\GeV$ using 3~\invab of $pp$ collisions at the HL-LHC.
Both conservative~(conserv.) and background-free (bg.-free) projections are shown, which bracket the expected \anubis sensitivity.
For comparison, the reach of other proposed transverse detectors \codex~\cite{CODEX-b:2019jve}, 
%MATHUSLA 200~m~\cite{Chou:2016lxi} 
and \mathusla 40~m~\cite{MATHUSLA:2025zyt} at the HL-LHC are also presented, alongside the latest LHC results from \atlas~\cite{ATLAS:2025pak} and \cms~\cite{CMS:2021juv} using $\sim140$~\invfb of $pp$ collisions.
}

\renewcommand{\arraystretch}{1.1}
\begin{tabular}{l|cc}
\hline
\hline
 & \textbf{$m_s=15~\text{GeV}$} & \textbf{$m_s=40~\text{GeV}$} \\
\hline
\textbf{\atlas, 140~\invfb} & \textcolor{black}{$[4.6 \times 10^{-1},2.7 \hspace{1cm}]$~m} & \textcolor{black}{$[1.3,\hspace{1.1cm} 4.6\hspace{1.cm}]$~m} \\
\textbf{\cms, 137~\invfb} & - & - \\
\textbf{\codex} & \textcolor{black}{$[2.5,\hspace{1.17cm} 6.1 \times 10^{1}]$~m} & - \\
\textbf{\mathusla 40} & \textcolor{black}{$[4.8,\hspace{1.17cm} 5.8 \times 10^{2}]$~m} & - \\
% \textbf{ANUBIS 1DV 5$\sigma$ conserv.} & \textcolor{black}{$[3.7 \times 10^{-1}, 1.5 \times 10^{2}]$~m} & \textcolor{black}{$[1 \times 10^{0}, 3.9 \times 10^{2}]$~m} \\
% \textbf{ANUBIS 1DV 5$\sigma$ bg.-free} & \textcolor{black}{$[1.1 \times 10^{-1}, 4 \times 10^{3}]$~m} & \textcolor{black}{$[3.2 \times 10^{-1}, 9.3 \times 10^{3}]$~m} \\
% \textbf{ANUBIS 2DVs 5$\sigma$ conserv.} & \textcolor{black}{$[8.6 \times 10^{-1}, 1.1 \times 10^{1}]$~m} & \textcolor{black}{$[2.6 \times 10^{0}, 3.4 \times 10^{1}]$~m} \\
% \textbf{ANUBIS 2DVs 5$\sigma$ bg.-free} & \textcolor{black}{$[2.7 \times 10^{-1}, 7.9 \times 10^{1}]$~m} & \textcolor{black}{$[7 \times 10^{-1}, 2.2 \times 10^{2}]$~m} \\
% \textbf{ANUBIS 1+2DVs 5$\sigma$ conserv.} & \textcolor{black}{$[2.7 \times 10^{-1}, 1.5 \times 10^{2}]$~m} & \textcolor{black}{$[7 \times 10^{-1}, 3.9 \times 10^{2}]$~m} \\
%\textbf{ANUBIS 1+2DVs 5$\sigma$ bg.-free} & \textcolor{black}{$[1.1 \times 10^{-1}, 4.0 \times 10^{3}]$~m} & \textcolor{black}{$[3.2 \times 10^{-1}, 9.3 \times 10^{3}]$~m} \\
\textbf{\anubis 1DV CLs conserv.} & \textcolor{black}{$[2.4 \times 10^{-1}, 4.3 \times 10^{2}]$~m} & \textcolor{black}{$[6.7 \times 10^{-1}, 1.2 \times 10^{3}]$~m} \\
\textbf{\anubis 1DV CLs bg.-free} & \textcolor{black}{$[1.1 \times 10^{-1}, 4.0 \times 10^{3}]$~m} & \textcolor{black}{$[3.2 \times 10^{-1}, 9.3 \times 10^{3}]$~m} \\
\textbf{\anubis 2DVs CLs conserv.} & \textcolor{black}{$[5.5 \times 10^{-1}, 2.2 \times 10^{1}]$~m} & \textcolor{black}{$[1.6,\hspace{1.17cm} 6.8 \times 10^{1}]$~m} \\
\textbf{\anubis 2DVs CLs bg.-free} & \textcolor{black}{$[2.7 \times 10^{-1}, 7.9 \times 10^{1}]$~m} & \textcolor{black}{$[7.0 \times 10^{-1}, 2.2 \times 10^{2}]$~m} \\
\textbf{\anubis 1+2DVs CLs conserv.} & \textcolor{black}{$[2.4 \times 10^{-1}, 4.3 \times 10^{2}]$~m} & \textcolor{black}{$[6.7 \times 10^{-1}, 1.2 \times 10^{3}]$~m} \\
\textbf{\anubis 1+2DVs CLs bg.-free} & \textcolor{black}{$[1.1 \times 10^{-1}, 4.0 \times 10^{3}]$~m} & \textcolor{black}{$[3.2 \times 10^{-1}, 9.3 \times 10^{3}]$~m} \\
\hline
\hline
\end{tabular}
\label{tab:ctau_reach_II}
\end{table*}

\begin{figure}[tb]
\centering
\begin{subfigure}[b]{0.45\textwidth}
\includegraphics[width=\textwidth]{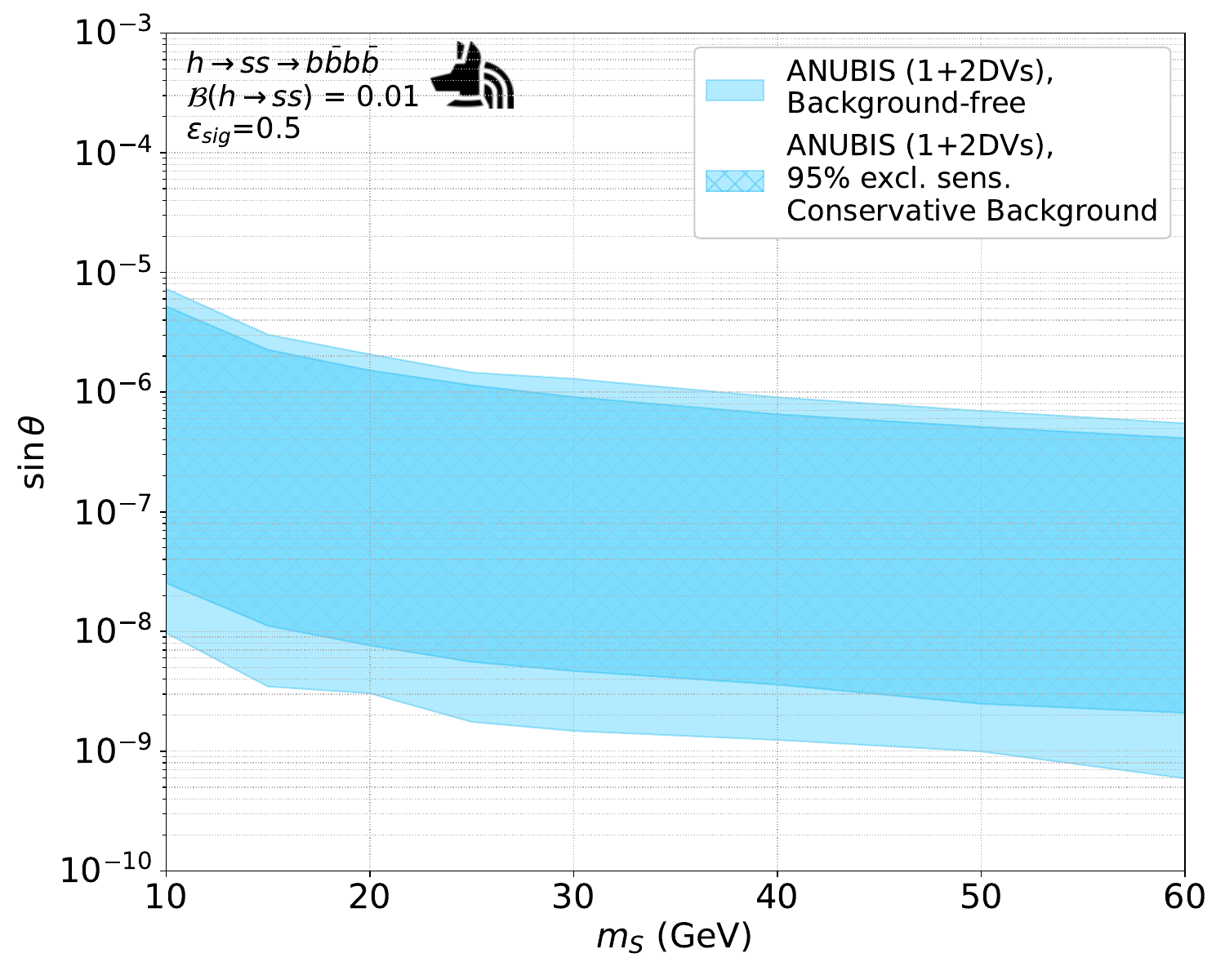}
\caption{}
\end{subfigure}
\begin{subfigure}[b]{0.45\textwidth}
\includegraphics[width=\textwidth]{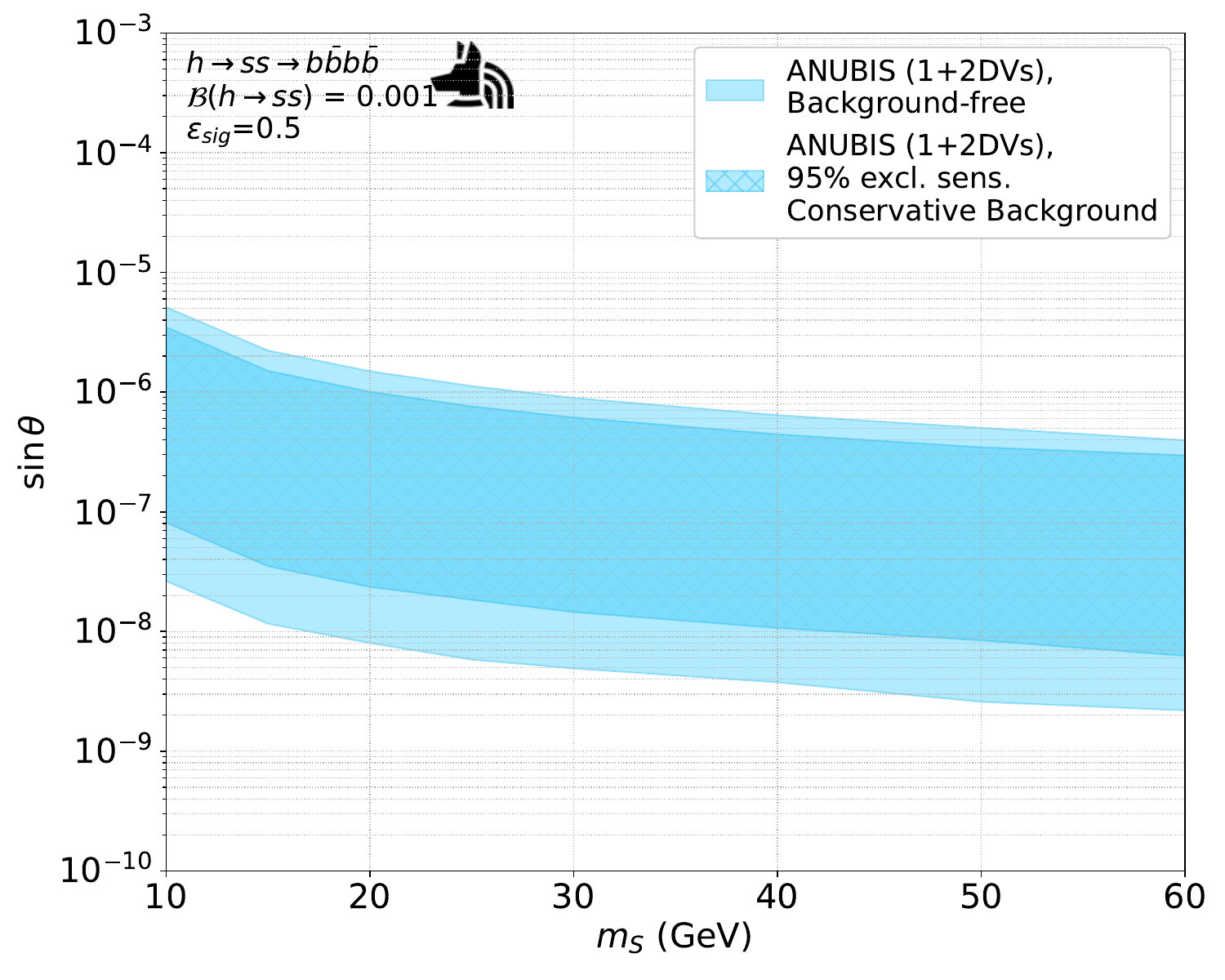}
\caption{}
\end{subfigure}
%\end{center}    
\caption{
\label{fig:HtoSS_Sensitivity_BC5}
The combined 1DV and 2DVs 95\% CL exclusion $\sin\theta$ reach of \anubis in the ceiling configuration for $h\to ss$ decays using 3~\invab of $pp$ collisions at the HL-LHC, where $s$ are long-lived scalars decaying as $s\to b\bar b$, and $\theta$ is the mixing angle between the $h$ and $s$ scalars. For (a) BC5 benchmark with $\br(h\to ss) = 10^{-2}$~\cite{Beacham:2019nyx}, and (b) for $\br(h\to ss) = 10^{-3}$.
Both conservative~(hatched area) and background-free~(solid filled area) projections are shown, which bracket the expected \anubis sensitivity.
}
\end{figure}

The signal model used to determine these sensitivity limits corresponds to the BC5 benchmark~\cite{Beacham:2019nyx} proposed by the CERN Physics Beyond Colliders (PBC) study group~\cite{Alemany:2019vsk}.
In BC5, the key parameter of interest is the mixing angle, \(\theta\), between the scalar LLP and the Higgs boson assuming a fixed $\br(h\to ss)$.
%rather than using this as a free parameter, as in Figures~\ref{fig:SensitivityAll} and~\ref{fig:SensitivityCeilingOnly}.
Hence, the above sensitivity projections can be translated into constraints on $\theta$ using the formalism from Ref.~\cite{Curtin:2013fra}, where this quantity is denoted by $\theta_s$. Additional mass points at $m_s=20$, 25, 30, and 50 GeV were also considered. 
The $\sin\theta$ interpretation is done assuming the BC5 recommendation of $\br(h\to ss)=0.01$, and the associated limits are shown in Figure~\ref{fig:HtoSS_Sensitivity_BC5}~(a) for the combined 1DV and 2DVs case. 
%However, since ANUBIS is capable of probing significantly lower branching ratios, this benchmark underestimates its full sensitivity.
To better represent \anubis' reach, the limits are also evaluated for \(\br(h\to ss) = 10^{-3}\) in Figure~\ref{fig:HtoSS_Sensitivity_BC5}~(b).
Remarkably, despite the branching ratio being reduced by one order of magnitude, the reach in $\sin\theta$ remains within approximately half-an-order of magnitude.

%%%%%%%%%%%%%%%%%%%%%%%%%%%%%%%%%%%5
\section{Conclusion and outlook} 
\label{sec:Summary}

In summary, the \anubis detector offers an unprecedented opportunity to dramatically extend the discovery potential of HL-LHC in the search for new physics beyond the SM through the detection of electrically neutral LLPs.
This is achieved by instrumenting a large decay volume adjacent to the \atlas experiment at the HL-LHC with tracking detectors.
Due to its location transverse to the beamline and close to the \atlas interaction point, \anubis will provide unique sensitivity to LLPs with decay lengths of $>$\order{10~\metre} produced at partonic centre-of-mass energies $\sqrt{\hat s}$ corresponding to the electroweak scale and above.
Hence, the \anubis detector will complement the discovery potential of the main HL-LHC detectors like \atlas, \cms, and LHCb and forward/fixed target detectors like FASER/SHiP, which are limited in their decay length and $\sqrt{\hat s}$ reach, respectively, cf.~Figure~\ref{fig:LLP_Complimentarity}.

The two proposed detector layouts for \anubis were outlined, their physics potential was assessed, and comparisons to other experiments were drawn.
The ceiling configuration of the \anubis detector was identified as the preferred one, both in terms of practical considerations for the installation and HL-LHC / \atlas operations, and in terms of the physics potential. 
The ceiling configuration is now the default for the \anubis  detector.
It foresees instrumenting the ceiling of the UX15 \atlas experimental cavern and the bottom of the PX14 and PX16 service shaft with tracking stations using  RPC detectors.

The potential sources of background were discussed, identifying hadronic interactions of $K_L^0$ and $n$ with material as the dominant contribution, which can be strongly reduced using dedicated selections.
This underlines the main advantage of the \anubis detector featuring a large air-filled decay volume between the \atlas detector and the cavern ceiling, combined with the adjacency to the interaction point.
The background contributions were estimated using a data-driven method. 
For the default ceiling configuration of the \anubis detector, an exclusion sensitivity at 95\% CL was found, corresponding to at most 36 signal events.
For a discovery sensitivity at 5$\sigma$ CL, 102 of exotic LLPs events were found to be sufficient under conservative assumptions, leaving ample potential for further background reduction.
The topology of potential background events was studied using MC simulations to guide future sensitivity optimisations.

The sensitivity of \anubis was evaluated in the context of the Benchmark Case 5 of the Physics Beyond Colliders initiative that features two scalar LLPs $s$ produced in an exotic decay of the 125~GeV Higgs boson and  decaying into $b\bar b$ quark pairs each.
Overall, \anubis is expected to probe $\br(h\to ss)$ down to $1.3\times 10^{-5}$.
This marks an improvement of over an order of magnitude beyond 
projected \atlas and \cms capabilities at the HL-LHC, and surpasses other transverse experiment proposals \codex and \mathusla.
Assuming further sensitivity improvements beyond the conservative assumptions in the background estimate, a sensitivity to $\br(h\to ss)$ of down to $3.3\times 10^{-6}$ could be achieved.
For $\br(h\to ss)=0.001$, an unprecedented $c\tau$ range from \order{0.1~\metre} to \order{10^3~\metre} could be probed according to conservative projections.
This reach could be potentially extended up to $c\tau \approx 10^4~\metre$.

\subsection*{Outlook} 
Active R\&D efforts aimed at the realisation of the \anubis detector are underway. 
The aforementioned prototype demonstrator \proanubis consisting of one \anubis tracking station module has recorded 177~\invfb of $pp$ collision data at $\sqrt s = 13.6~\TeV$ in 2024--2025~\cite{ANUBIS:2025mme,ANUBIS:2025inn}. 
The analysis of \proanubis data is ongoing, allowing to refine the tracking and vertexing algorithms, inform detector performance benchmarks~\cite{ANUBIS:2026jxg}, improve background simulations, and further geometry optimisations for \anubis.
Furthermore, strong and concerted efforts are underway to systematically study alternative RPC gas mixtures within the context of the DRD1 Collaboration~\cite{Shah:2025luk,Ahmadi:2025pjv,RPCEcoGasGIF:2025gqp}. 
The goal is adhering to EU regulations by achieving a global warming potential below 150 times that of CO$_2$ without compromising on detector performance.
Several promising mixtures that are expected to perform well in the moderate rate environment $\gtrsim$20~m away from the \atlas IP have been identified. 
In parallel, efforts are underway to assess and ensure the mechanical integrity of RPC gas gaps to prevent any gas leaks at the source. 
A novel methodology for accelerated ageing of RPC gas gaps using thermal cycling has been established in the context of the Phase~II upgrade of the \atlas RPCs, informing the re-design of their gas gaps~\cite{Shah:2025pcy}.

Beyond this work, concerted efforts are underway to map out the sensitivity of the \anubis detector to alternative LLP scenarios. 
To facilitate these studies, a user-friendly modular framework \setanubis has been developed.
The primary focus is to map out the sensitivity to PBC benchmark models. 
For example, the sensitivity to heavy neutral leptons (HNLs) that couple exclusively to first or second generation leptons, i.e., PBC Benchmark Cases 6 and 7, has been published recently~\cite{Bauer:2026ivt}. 
Beyond PBC benchmarks, other scenarios where \anubis + \atlas are expected to provide unprecedented sensitivity such as photophobic axion-like particles produced in association with $W$ or $Z$ bosons are being actively explored.

\begin{acknowledgments}

The authors would like to express their gratitude to: the \atlas Collaboration for hosting \proanubis, facilitating the data taking and data analysis, and provision of consumables; the \atlas RPC community for their guidance with constructing the RPCs for \proanubis and the chamber cages, and maintenance/troubleshooting; and the \atlas Technical Coordination team for help with installing, connection of services, maintenance, and safety; to the colleagues in the \codex collaboration for lending front-end board electronics to allow for a timely installation of the \proanubis demonstrator; to the PBC BSM group and especially group conveners Felix Kahlhoefer and Torben Ferber for their guidance and efficient moderation of the scientific discussions, and finally, to colleagues in the wider LLP community for the scrutiny of our results and thoughtful suggestions. Particular thanks go to: Ben Allanach, Cristiano Alpigiani, David Curtin, Albert De Roeck, Miriam Diamond, Erez Etzion, Simon Knapen, Gaurav Kumar, Steven Lowette, Maxim Ovchinnikov, Runze Ren, Heather Russell, Richard Shaw, and Gustavo Uribe.
We gratefully acknowledge the support of UKRI under the Future Leaders Fellowship scheme.
\end{acknowledgments}

\bibliographystyle{JHEP}
\bibliography{bibs/main,bibs/anubis-standard-refs}

\end{document}